\documentclass[aps,prb]{revtex4}
\usepackage{graphicx,amssymb,amsfonts,amsmath}

\begin{document}

\title{$\frac{1}{N}$ expansion of the nonequilibrium infinite-U Anderson Model}
\author{Zurab Ratiani and Aditi Mitra}
\affiliation{Department of Physics, New York University, 4
Washington Place, New York, New York 10003}
\date{\today}

\begin{abstract}
Results are presented for the nonequilibrium infinite-U Anderson model using a large-N approach, where
$N$ is the degeneracy of the impurity level, and where nonequilibrium is established by
coupling the level to two leads at two different chemical potentials so that there is current flow. 
A slave-boson representation combined with Keldysh functional
integral methods is employed. 
Expressions for the static spin susceptibility $\chi_S$ and the conductance $G$ are 
presented to ${\cal O}\left(\frac{1}{N}\right)$ and for an applied voltage difference $V$ less than the 
Kondo temperature. The correlation function for the slave-boson is found to be significantly modified from
its equilibrium form in that it acquires a rapid decay in time with a rate 
that equals the current induced decoherence rate.
Physical observables are found to have a rather complex dependence on 
the coupling strength to the two leads which can lead to
asymmetric behavior $\chi_S(V) \neq \chi_S(-V)$, 
$G(V)\neq G(-V)$ both in the mixed valence and in the Kondo regime.

\end{abstract}


\maketitle

\section{Introduction}

The theoretical problem of strong correlations coupled with nonequilibrium has become
an active area of research in recent years, in part due to the 
enormous success in 
realizing experimental systems which can be driven out of equilibrium in a controlled manner. 
Some examples of these are  current carrying
quantum dots and single molecule devices~\cite{nanoexp}, 
strongly driven ferromagnetic systems~\cite{fmrexp}, 
cold atoms trapped in optical lattices with rapidly tunable parameters~\cite{coldatomexp}. 
One of the theoretical challenges in the study of out of equilibrium strongly correlated systems is that, 
unlike systems in equilibrium which are characterized by some underlying principles 
such as the energy minimization principle, 
no basic underlying principles are known for out of  
equilibrium systems making it rather difficult to develop general theoretical
techniques to study them. 

Perhaps the most actively studied out of equilibrium systems are nonequilibrium 
quantum impurity models which are systems characterized by a few local
degrees of freedom coupled to one or more reservoirs (as in a quantum dot or a molecular conductor), 
and where nonequilibrium is
achieved by
maintaining the reservoirs at different chemical potentials and/or by subjecting the system to 
time-dependent fields.
For strong local
interactions the ground state of quantum impurity models show
many-body resonances such as the Kondo or polaronic resonance. The
effect of current flow on these resonances has been studied using a variety 
of methods such as renormalized
perturbation theory~\cite{Paaske04}, flow equation methods~\cite{Kehrein05}, 
real time renormalization group on the Keldysh contour~\cite{Schoeller07,Mitra07}, 
and functional renormalization group methods~\cite{Meden07}. 
While these approaches are applicable when the external drive is large
as compared to the Kondo temperature, in the opposite limit of drive small compared
to the Kondo temperature, 
perturbative methods based on Fermi-liquid theory have been used~\cite{neqfermiliquid}.
There have also been efforts at developing exact solutions based on the construction of
exact scattering states in the presence of current flow~\cite{Ludwig02,Andrei06}.
There are also several promising numerical methods that 
are being developed such as the real-time numerical renormalization group method~\cite{Anders08},
quantum Monte Carlo computation of real time Keldysh diagrams~\cite{Rabani07,Werner08}, 
iterative summation of real time path integrals~\cite{Egger08} and the imaginary time formulation of
real-time nonequilibrium problems~\cite{Han07}.

In this paper we will use large-N methods~\cite{Bickers87} to
study a nonequilibrium quantum impurity model.   
In particular, we will study the Anderson model when the on-site Coulomb interaction $U=\infty$, and 
in addition
the system has been driven out of equilibrium due to current flow. $N$ here will represent the degeneracy
of the impurity level. The physical systems this corresponds to are quantum dots
or molecular devices where the level active in transport is characterized by a total angular momentum
$J = L + S$ which is large, and hence has a large degeneracy $N = 2J + 1$. 
This could arise due to the particular form of the confining potential in
the quantum dot, or by the use of a molecule where conduction occurs via a metal ion with a 
partially empty 
outermost $d$ or $f$ orbital. Note that the infinite-U Anderson model under out of equilibrium 
conditions has so far been 
studied using the non-crossing approximation (NCA)~\cite{neqNCA} and slave boson mean-field methods
~\cite{neqslavebosonmf}. In this paper we will also employ the slave-boson representation which is
a convenient way to project out all states except the empty and singly
occupied state of the dot~\cite{ColemanBarnes}. However,  
we will go beyond mean-field by including the effect
of fluctuations to ${\cal O}\left(\frac{1}{N}\right)$. Our theoretical approach is closest to that of
Read {\it et al.}~\cite{Read85,Read87}, but carried out for a nonequilibrium system using 
Keldysh functional
integral methods. 

A few words on the regime of validity of the results presented in this
paper. The $U=\infty$ limit of the Anderson model 
is the so called mixed-valence regime where the system is characterized
by both local charge as well as spin fluctuations.  
The Kondo regime may be accessed
by making the bare level energy large and negative in which case the 
charge fluctuations are frozen out and only the 
spin-fluctuations exist. This limit can be taken in a straightforward way 
in all physical observables. Thus we will present results for the  nonequilibrium 
static spin susceptibility and the conductance in both the mixed valence
as well as in the Kondo regime.  

This paper is organized as follows. 
In section~\ref{model} we present the model, and write it as a 
Keldysh path-integral suitable for studying nonequilibrium systems. In section~\ref{results} we briefly
present the main results of the paper before turning to the full calculation. 
In section~\ref{meanfield} we study
the Keldysh path integral in the limit of $N\rightarrow\infty$ when a mean-field or 
saddle-point approximation
becomes exact. In this limit the voltage dependence of the local charge density, static
susceptibility and the conductance are derived.    
Following this, the rest of the paper is devoted to the study of the
effect of fluctuations to
${\cal O}\left(1/N\right)$.  As found by Read et al~\cite{Read85}, the $1/N$ corrections are 
in general associated with
infra-red divergences whose origin is the zero-mode of the slave-boson representation.  
While the infrared divergences are logarithmic in equilibrium, we find that out of equilibrium 
the divergences
become more severe with a pole structure. 
However, just as in 
equilibrium, in the computation of all physical observables these infrared
divergences are found to cancel 
so that the final expressions are well defined. 
  
The ${\cal O}(1/N)$ computation is organized as follows. 
In section~\ref{fluc} the mean-field saddle point expressions for the 
level position and the level broadening are corrected to ${\cal O}(1/N)$. 
In Section~\ref{nF} the local impurity charge density is computed. In Section~\ref{Xray}
the bosonic correlation function is evaluated. While in equilibrium the bosonic correlation
function has a power-law decay in time with an
exponent consistent with X-ray edge physics~\cite{Read85}, for the current carrying case we find 
that the long time behavior has both a power-law as well as
a rapid exponential decay in time, the latter arising due to current induced decoherence. The bosonic
correlation function appears in the computation of various physical observables.   
We present results for the static
susceptibility in section~\ref{susc}, while expressions for the impurity spectral density and 
conductance are presented in~\ref{spect}. Many of the details
of the computation are relegated to the appendices. Finally we conclude in section~\ref{conc}.

\section{Model} \label{model}

We use the slave-boson representation~\cite{Bickers87} of the infinite-U Anderson model which
is a convenient way to project out all except the empty and singly occupied states of the impurity level.
The Hamiltonian in this representation is, 
\begin{eqnarray}
H = \sum_m E_0 f_m^{\dagger} f_m + \sum_{k,m,\alpha} \epsilon_k c_{km\alpha}^{\dagger} c_{km\alpha} +
\frac{1}{\sqrt{N}}\sum_{k,m,\alpha=L,R} V_{\alpha}\left(c_{km\alpha}^{\dagger}f_m b^{\dagger}
+ f_m^{\dagger}c_{km\alpha}b\right) \label{H}
\end{eqnarray}
where $m = -J \ldots J$ represents the spin-projection of the local level, N=2J+1 is
the degeneracy of the level, $c_{km\alpha}$ 
represent the lead electrons, and     
we have generalized to the case where there are
two leads (labeled by $\alpha=L,R$) which will be maintained at two
different chemical potentials $\mu_{L,R}$ to capture the nonequilibrium current carrying case.
$\frac{V_{\alpha=L,R}}{\sqrt{N}}$ is the hybridization to the two the leads. 
The above Hamiltonian is accompanied by the constraint
\begin{equation}
1 = \sum_m f_m^{\dagger} f_m +  b^{\dagger} b \label{const}
\end{equation}
to ensure that the system remains within the restricted Hilbert space of an empty or singly 
occupied local level. 

We write the Keldysh path integral~\cite{Kamenev04} 
for Eq.~\ref{H} and impose the constraint in Eq.~\ref{const} by
introducing two Lagrange multipliers $\lambda_{\pm}$
\begin{eqnarray}
Z_K = \int {\cal D}\left[f_{m\pm},\bar{f}_{m\pm},\lambda_{\pm},b_{\pm},b^*_{\pm},c_{m},\bar{c}_m\right]
\exp{\left(i Tr[S_K]\right)} \label{action1}
\end{eqnarray}
where the Tr symbol in Eq.~\ref{action1} represents a trace over time indices, and 
\begin{eqnarray}
&&S_K = \sum_m \begin{pmatrix} \bar{f}_{m-}& \bar{f}_{m+} \end{pmatrix}
\begin{pmatrix}i\partial_t -E_0  &0 \\
0 & i\partial_t - E_0  \end{pmatrix} \begin{pmatrix} f_{m-}\\ f_{m+}\end{pmatrix} + \sum_{km\alpha}
\begin{pmatrix} \bar{c}_{km\alpha -} & \bar{c}_{km\alpha +} \end{pmatrix}
g_{c\alpha}^{-1} \begin{pmatrix} c_{km\alpha -}\\ c_{km\alpha +}\end{pmatrix} \nonumber \\
&& 
 + \sum_{km\alpha}
\begin{pmatrix} \bar{f}_{m-} & \bar{f}_{m+} \end{pmatrix}
\begin{pmatrix} -\frac{V_{\alpha}}{\sqrt{N}} b_- & 0 \\ 0 & 
\frac{V_{\alpha}}{\sqrt{N}} b_+\end{pmatrix} \begin{pmatrix} c_{km-}\\ c_{km+}\end{pmatrix}
+ \sum_{km\alpha}
\begin{pmatrix} \bar{c}_{km-} & \bar{c}_{km+} \end{pmatrix}
\begin{pmatrix} -\frac{V_{\alpha}}{\sqrt{N}} b_{-}^* & 0 \\ 0 & \frac{V_{\alpha}}{\sqrt{N}} b_{+}^* \end{pmatrix} \begin{pmatrix} f_{m-}\\ f_{m+}\end{pmatrix}
\nonumber\\
&& + \begin{pmatrix} {b}^*_{-} & {b}^*_{+} \end{pmatrix}
\begin{pmatrix} i\partial_t & 0 \\ 0 & i \partial_t \end{pmatrix}
\begin{pmatrix} b_{-}\\ b_{+}\end{pmatrix} - \lambda_-\left[\sum_m \left(\bar{f}_{m-}f_{m-} + \frac{1}{2}\right)+ b^{*}_{-} b_- -1 \right] + \lambda_+
\left[\sum_m \left(\bar{f}_{m+}f_{m+}+ \frac{1}{2}\right)+ b^{*}_+ b_+ -1 \right]\label{SK1}
\end{eqnarray}
In the above $g_{c\alpha}^{-1}$ is the inverse Green's function for the leads and is a $2\times2$ matrix in
Keldysh space. It is convenient to integrate out the lead electrons to obtain,
\begin{eqnarray}
&&S_K = \sum_m \begin{pmatrix} \bar{f}_{m-}& \bar{f}_{m+} \end{pmatrix}
\begin{pmatrix}i\partial_t -E_0 -\lambda_{-}  &0 \\
0 & i\partial_t - E_0 + \lambda_+ \end{pmatrix} \begin{pmatrix} f_{m-}\\ f_{m+}\end{pmatrix}
\nonumber \\
&& + \begin{pmatrix} {b}^*_{-} & {b}^*_{+} \end{pmatrix}
\begin{pmatrix} i\partial_t-\lambda_-  
& 0 \\ 0 & i \partial_t +\lambda_+  \end{pmatrix}
\begin{pmatrix} b_{-}\\ b_{+}\end{pmatrix} \label{SK2}\\
&& - \frac{1}{N}\sum_{m\alpha } V_{\alpha}^2
\begin{pmatrix} \bar{f}_{m-} & \bar{f}_{m+} \end{pmatrix}
\begin{pmatrix} b_- & 0 \\ 0 & -b_+\end{pmatrix}
\begin{pmatrix} g^{c,\alpha}_{--} & g^{c,\alpha}_{-+} \\ g^{c,\alpha}_{+-} & g^{c,\alpha}_{++}\end{pmatrix}
\begin{pmatrix} b^*_- & 0 \\ 0 & -b^*_+\end{pmatrix}
\begin{pmatrix} f_{m-}\\ f_{m+}\end{pmatrix}
+ \left(\lambda_- - \lambda_+ \right)\left(1 - \frac{N}{2} \right)\nonumber
\end{eqnarray}

Performing a rotation
to retarded (R), advanced (A), Keldysh (K) space~\cite{Kamenev04}, and defining 
the quantum fields as $O_q = (O_--O_+)/2$ and the classical field as $O_{cl}= (O_- + O_+)/2$, we get
\begin{eqnarray}
&&S_K = \sum_m \begin{pmatrix} \bar{f}_{m,q}& \bar{f}_{m,cl} \end{pmatrix}
\left[g_{0f}^{-1} - \lambda_{cl}\tau_0 - \lambda_q \tau_x -
\left(b_{cl}\tau_0 + b_q \tau_x \right) \Sigma_c \left(b^*_{cl}\tau_0 + b^*_q \tau_x \right) \right]
\begin{pmatrix} f_{m,cl}\\ f_{m,q}\end{pmatrix}
\nonumber \\
&& + 2\begin{pmatrix} {b}^*_{cl} & {b}^*_{q} \end{pmatrix}
\begin{pmatrix}  -\lambda_q
&i\partial_t - \lambda_{cl}\\i \partial_t -\lambda_{cl} &  -\lambda_q \end{pmatrix}
\begin{pmatrix} b_{cl}\\ b_{q}\end{pmatrix} + 2 \lambda_q\left(1 - \frac{N}{2} \right)\label{SK3}
\end{eqnarray}
where the $\Sigma_c$ are the self-energies due to coupling to leads,
\begin{eqnarray}
\Sigma_c = \begin{pmatrix} \Sigma^R_c & \Sigma^K_c \\0 & \Sigma^A_c \end{pmatrix}
\end{eqnarray}
with
$\Sigma^{i=R,A,K}_c(t,t^{\prime}) = \frac{1}{N}\sum_{k,\alpha=L,R} V_{\alpha}^2 
g^{i=R,A,K}_c(k;t,t^{\prime})$. Thus
the self-energies due to coupling to leads is ${\cal O}\left(\frac{1}{N}\right)$. We will make the
assumption of constant density of states in the leads which gives  
\begin{eqnarray}
\Sigma^R_c =-\frac{i}{N}\pi \rho \sum_{\alpha=L,R} V_{\alpha}^2 
=-i\left(\Gamma_{L} + \Gamma_R\right) = -i \Gamma \label{gamm1}\\
\Sigma^A_c = i \Gamma \\
\Sigma^K_c = -2i \Gamma \sum_{\alpha=L,R}\frac{\Gamma_{\alpha}}{\Gamma}
\left(1- 2f(\omega -\mu_{\alpha})\right)
\end{eqnarray}

The aim will be to use the action in Eq.~\ref{SK2} to 
evaluate physical observables perturbatively in $1/N$. Before turning to
the full computation, we present the main results in the next section.

\section{Brief discussion of results}~\label{results}

Let us suppose that the chemical potential of the left lead is $\mu_L = V/2$ while that
of the right lead is $\mu_R = -V/2$. As specified in
Eq~\ref{gamm1}, let $\Gamma_{L}(\Gamma_R)$ be the self-energy due to coupling to the
left (right) lead, while $\Gamma = \Gamma_L + \Gamma_R$ is the total self-energy. 
In terms of the above parameters, 
the static susceptibility in the Kondo regime (denoted
by the superscript $n_F=1$ to indicate the value of the charge on the impurity level) is
found to have the following universal form,
\begin{eqnarray}
&&\chi_S^{n_F=1} = \frac{g^2\mu_B^2 J(J+1)}{3 T_K}
\left[1 + 1.5 \frac{\Gamma_L \Gamma_R}{\Gamma^2}\left(\frac{V}{T_K}\right)^2
+\frac{1}{N}\left(\frac{\Gamma_L -\Gamma_R}{\Gamma}\right)\left(\frac{V}{2T_K}\right) C_{S1} +
\right.\nonumber \\ 
&&\left. \frac{1}{N}\left(\frac{V}{2T_K}\right)^2 \left(C_{S2} + C_{S3} -C_{S1}\right)  
-\frac{1}{N}\left(4.5+3C_{S0} + C_{S3}-C_{S1}\right)\frac{\Gamma_L\Gamma_R}{\Gamma^2}\left(\frac{V}{T_K}\right)^2 
\right] \label{suscsumm}
\end{eqnarray}
where $T_K = T_A^0\left(1 - \frac{C_{S0}}{N}\right)$ is the Kondo temperature
correct to ${\cal O}\left(1/N\right)$, with $T_A^0= D e^{\frac{\pi E_0}{N\Gamma}} $ 
being the mean-field Kondo temperature
~\cite{Read87},
and the $C_{Si}$ are numbers specified in
the text (after Eq.~\ref{neqsusc3}).  
Thus one finds that for an asymmetric coupling to leads ($\Gamma_L \neq \Gamma_R$), 
$\chi_S(V) \neq \chi_S(-V)$. This lack of symmetry 
when $V\leftrightarrow -V$
arises due to the fermi-level
dependence of the Kondo temperature.
To see this we set the coupling to one 
of the leads (say $\Gamma_R$) to zero. This 
corresponds to an equilibrium configuration 
where there is no current flow. For this case Eq.~\ref{suscsumm}
reduces to
\begin{eqnarray}
&&\chi_S^{n_F=1}(\mu_L = V/2,\Gamma_R=0) = \frac{g^2\mu_B^2 J(J+1)}{3 T_K}
\left[1 + \frac{1}{N}\left(\frac{V}{2T_K}\right)C_{S1}
+\frac{1}{N}\left(\frac{V}{2T_K}\right)^2\left(C_{S2} + C_{S3} - C_{S1}\right) \right]
\label{suscsummeq}
\end{eqnarray}
Thus the terms 
in Eq.~\ref{suscsummeq} can be interpreted as a change in the Kondo temperature arising from 
a change in the chemical potential of the left lead by $\delta \mu_L = V/2$.  The 
asymmetry $\chi_S(V) \neq \chi_S(-V)$ in the Kondo regime therefore arises 
when the level is unequally coupled to two leads, each associated with a different equilibrium
Kondo temperature. 
In contrast, the terms of the type
$\frac{\Gamma_L\Gamma_R}{\Gamma^2}\left(\frac{V}{T_K}\right)^2$ in Eq~\ref{suscsumm}
are purely nonequilibrium
terms that arise due to inelastic scattering processes in the energy window
$V$ when there is
current flow, and are thus associated with current induced decoherence. 
The identification of these terms with
decoherence becomes clearer below when we discuss the slave-boson correlation function.

We now turn to the discussion of the conductance. Here too one finds that the fermi-level dependence of
the spectral density can give rise to a 
conductance that is asymmetric under $V\rightarrow -V$~\cite{Affleck08}.
In particular the mean-field saddle point expression 
for the conductance in the mixed valence regime is found to be  
\begin{eqnarray}
G_{sp}(V)= G_{sp}(V=0)\left[1- 
\left(\frac{\Gamma_L -\Gamma_R}{\Gamma}\right)\left(\frac{2V}{T_A^0}\right)
-\frac{12\Gamma_L \Gamma_R}{\Gamma^2}\left(\frac{V}{T_A^0}\right)^2 +3
\left(\frac{V}{T_A^0}\right)^2 - 3n_F\frac{\Gamma_L \Gamma_R}{\Gamma^2} 
\left(\frac{V}{T_A^0}\right)^2\right]\label{Gspsumm}
\end{eqnarray}
where $n_F$ is the charge density on the level when $\mu_L =\mu_R=0$, and $G_{sp}(V=0)=
\frac{N e^2}{h}\frac{4\Gamma_L \Gamma_R}{\Gamma^2} \left(\frac{\pi n_F}{N}\right)^2$. The
conductance in the Kondo regime can be accessed by taking the limit $n_F \rightarrow 1$
in Eq.~\ref{Gspsumm}.
Thus for a symmetric coupling to the two leads, the mean-field conductance in the Kondo regime becomes 
\begin{eqnarray}
G^{n_F=1}_{sp}(V;\Gamma_L=\Gamma_R)= G^{n_F=1}_{sp}(V=0)\left[1 - \frac{3}{4} 
\left(\frac{V}{T_A^0}\right)^2\right]\label{GspsummK}
\end{eqnarray}
The
$1/N$ correction to the conductance for the case of symmetric couplings to leads is given
in Eq.~\ref{G1Nneq} for the mixed valence regime and in Eq.~\ref{GneqNK} in the Kondo regime. 

We now turn to the discussion of the bosonic correlation function 
$\bar{D}_K(t,t^{\prime}) = -i\langle \{b(t),b^{\dagger}({t^{\prime}})\}\rangle$ which is
used to obtain the physical observables discussed above. At the mean-field level,
$b(t) \rightarrow \langle b\rangle$ in the Hamiltonian 
(Eq.~\ref{H}), so that the $U(1)$ symmetry of the Hamiltonian 
is broken.  
In equilibrium, including fluctuations to ${\cal O}(1/N)$, the
correlation function becomes~\cite{Read85,Read87} 
\begin{eqnarray}
\bar{D}^K_{eq}(t) = -2i\left(1-n_F\right)\left(1-\frac{n_F^2}{N}\ln\left(tT_A^0\right)\right)
\label{DKeq1}
\end{eqnarray}
It was argued that~\cite{Read85,Read87} since the
model in Eq.~\ref{H} cannot have any broken symmetry state, 
including terms to higher orders in $1/N$ should
lead to a power-law decay in the bosonic correlation function so that the symmetry 
of the Hamiltonian is restored. 
Thus,  
\begin{equation}
\bar{D}^K_{eq}(t) \sim \frac{1}{\left(tT_A^0\right)^{\alpha}} \label{DKeq2}
\end{equation}
where $\alpha = N\frac{\delta^2}{\pi^2} = \frac{n_F^2}{N}$ and 
equals the Nozieres-de Dominicis infrared exponent for the response of an electron gas subjected
to a sudden change in potential~\cite{nozieres}. 

We find that the result for the bosonic correlation function for the current carrying case, 
and for long times ($Vt \gg 1 $) is,
\begin{eqnarray}
\bar{D}^K_{neq}(t) = -2i\left(1-n_F\right)\left(1-\frac{c_L + c_R}{N}\ln\left(tT_A^0\right)
- \frac{c_{dec}}{N} V t\right)
\label{DKneq1}
\end{eqnarray}
where the coefficients $c_{L,R,dec}$ are specified in Eqns.~\ref{cL},~\ref{cdec}. The $c_i$ are weakly 
voltage dependent, and neglecting terms of ${\cal O}\left(V/T_A^0\right)$ are,
\begin{eqnarray}
c_{L,R} = n_F^2\frac{\Gamma_{L,R}^2}{\Gamma^2} + {\cal O}\left(\frac{V}{T_A^0}\right)\\
c_{dec} =  n_F^2\frac{2\Gamma_{L}\Gamma_{R}}{\Gamma^2} + {\cal O}\left(\frac{V}{T_A^0}\right)
\end{eqnarray}
For evaluating quantities to ${\cal O}(1/N)$, Eq.~\ref{DKneq1} is sufficient. However it is
interesting to consider how $\bar{D}^K$ would change when
higher order in $1/N$ terms are included. Following Eq.~\ref{DKeq2}, we expect that the bosonic
correlation function will have the form
\begin{equation}
\bar{D}^K_{neq}(t) \sim \frac{1}{\left(tT_A^0\right)^{\alpha_{neq}}}
\exp\left(- \frac{n_F^2}{N}\frac{2\Gamma_{L}\Gamma_{R}}{\Gamma^2} Vt\right) \label{DKneq2}
\end{equation}
where $\alpha_{neq} = (c_L + c_R)/N$. Thus to all orders in $1/N$
the bosonic correlation function will be characterized
with a long time power-law decay along with rapid exponential decay in time,
the latter arising due to current induced decoherence. The rate of decoherence is 
$\frac{\Gamma_L \Gamma_R}{\Gamma^2}V$, and is a energy scale that appears repeatedly in all
physical observables.
Note that 
Eq.~\ref{DKneq2} is also consistent with 
nonequilibrium X-ray edge physics {\sl i.e.}, the response of an out of equilibrium
electron gas to a sudden change in potential studied recently in various contexts~\cite{Ng}. 

We now turn to the derivation of the above results. 

\section{Mean-field saddle point treatment} \label{meanfield}

In the mean field saddle point treatment, one assumes the fields 
$b_{cl,q},\lambda_{cl,q}$ in Eq.~\ref{SK3} to be 
constants in
time. The action $S_K$ is then minimized both with respect to the classical
fields $b_{cl},\lambda_{cl}$ and the quantum fields $b_q, \lambda_q$. The
classical saddle points $\frac{\delta S_K}{\delta \lambda_{cl}}=0,\frac{\delta S_K}{\delta b_{cl}}=0 $ 
are automatically satisfied for $b_q =\lambda_q = 0$. Thus in order to satisfy 
the saddle point equations with respect to the quantum fields $\frac{\delta S_K}{\delta \lambda_{q}}=0,
\frac{\delta S_K}{\delta b_{q}}=0$ it is sufficient to
expand $S_K$ to linear order in
the quantum field.  To carry these steps out, we integrate out the fermionic fields in Eq.~\ref{SK3} to obtain
\begin{eqnarray}
&&S_K =
-iN Tr\ln \left[G_{mf}^{-1} - \lambda_q \tau_x - b_{cl}\Sigma_c b_q \tau_x - b_q \tau_x \Sigma_c b_{cl} +
O(b_q^2)\right] \label{SK11a}\\
&&+ 2 \lambda_q\left(1 -\frac{N}{2} \right)- 2 \lambda_q \left(b_q^2 + b_{cl}^2 \right) - 4 \lambda_{cl} b_{cl} b_q
\nonumber
\end{eqnarray}
where the mean-field fermionic Green's function is 
\begin{equation}
G_{mf}^{-1} = g^{-1}_{0f} - \lambda_{cl} - b_{cl}^2\Sigma_c \label{gmfdef}
\end{equation}
Defining 
\begin{equation}
\tilde{\Gamma} = b_{cl}^2 \Gamma
\end{equation}
where $\tilde{\Gamma}$ plays the role of the level broadening. 
\begin{eqnarray}
G^R_{mf}(\omega) = \frac{1}{\omega - E_0 -\lambda_{cl} + i \tilde{\Gamma}}\label{GRmfdef}\\
G^K_{mf}(\omega) = \left(\frac{\tilde{\Gamma}}{\Gamma}\right) G^R_{mf}(\omega) \Sigma^K_c(\omega) G^A_{mf}(\omega)
\label{GKmfdef}
\end{eqnarray}
From Eq.~\ref{SK11a}, the saddle point equation for $\lambda_q$, $\frac{\delta S_K}{\delta \lambda_q}=0$ gives
\begin{eqnarray}
2\left(1-\frac{N}{2} \right)
-  2 b_{cl}^2 - iN\left[\frac{\partial}{\partial \lambda_q}Tr\ln\left(G_{mf}^{-1} - \lambda_q \tau_x \right)
\right]_{\lambda_q=0}=0
\end{eqnarray}
This leads to,
\begin{eqnarray}
1-\frac{N}{2} = b_{cl}^2 - \frac{i N}{2}Tr\left[G^K_{mf}\right]
\label{splam}
\end{eqnarray}
Using Eq.~\ref{GKmfdef} the above becomes,
\begin{eqnarray}
1 = b_{cl}^2 + N \int \frac{d \omega}{2\pi} \frac{2 \Gamma b^2_{cl}}{(\omega - E_0 - \lambda_{cl})^2 + \Gamma b_{cl}^2}
\left( \frac{\Gamma_L}{\Gamma} f(\omega - \mu_L) + \frac{\Gamma_R}{\Gamma} f(\omega - \mu_R)\right)
\label{sp1}
\end{eqnarray}
After performing the frequency integrations, we obtain
\begin{eqnarray}
1 = \frac{\tilde{\Gamma}}{\Gamma} + \frac{N}{\pi} \left[\frac{\Gamma_L}{\Gamma}\arctan\frac{\tilde{\Gamma}}
{E_0 + \lambda_{cl} - \mu_L} + \frac{\Gamma_R}{\Gamma}\arctan\frac{\tilde{\Gamma}}{E_0 + \lambda_{cl} - \mu_R}
\right] \label{sp11}
\end{eqnarray}

Similarly, minimizing Eq.~\ref{SK11a} with respect to $b_{q,cl}$ leads to
\begin{eqnarray}
-4 \lambda_{cl} b_{cl} -i N \left[\frac{\partial}{\partial b_q} Tr \ln\left(G_{mf}^{-1}
- b_{cl}\Sigma_c b_q \tau_x - b_q \tau_x \Sigma_c b_{cl}\right)\right]_{b_q=0}=0
\label{spR}
\end{eqnarray}
Using expressions for $\Sigma_c$, the above leads to
\begin{eqnarray}
\frac{\lambda_{cl}}{\Gamma} + N\int \frac{d\omega}{2\pi} \left(G^R_{mf}(\omega) + G^A_{mf}(\omega) \right)
\left( \frac{\Gamma_L}{\Gamma} f(\omega - \mu_L) + \frac{\Gamma_R}{\Gamma} f(\omega - \mu_R)\right)=0
\label{sp2}
\end{eqnarray}
which after performing the frequency integrations gives,
\begin{eqnarray}
\frac{\lambda_{cl}}{\Gamma} + \frac{N}{\pi}\left[ \frac{\Gamma_L}{\Gamma}
\ln \frac{\sqrt{(\mu_L - E_0 - \lambda_{cl})^2 + \tilde{\Gamma}^2}}{D}
+ \frac{\Gamma_R}{\Gamma}
\ln \frac{\sqrt{(\mu_R - E_0- \lambda_{cl})^2 + \tilde{\Gamma}^2}}{D}
\right] =0 \label{sp22}
\end{eqnarray}
We will now proceed to solve the two saddle point equations Eq.~\ref{sp11} and ~\ref{sp22}, and use the solution
to evaluate various observables. The results obtained will be exact in the 
limit $N \rightarrow \infty$.

{\bf Solution of the saddle point equations}\newline
Let us define 
\begin{equation}
\lambda_{cl} + E_0 = \epsilon_F \label{eFdef}
\end{equation}
where $\epsilon_F$ is the effective position of the impurity level. 
When $N \rightarrow \infty$,  $\Gamma,\tilde{\Gamma}\rightarrow 0$ and $N \Gamma = const$. Using this, 
Eq.~\ref{sp11} may be simplified to
\begin{eqnarray}
1 = \frac{\tilde{\Gamma}}{\Gamma}  + \frac{N \tilde{\Gamma}}{\pi}\left[
\frac{{\Gamma}_L/\Gamma}{\epsilon_F - \mu_L} + \frac{\Gamma_R/\Gamma}{\epsilon_F - \mu_R}\right] \label{sp11a}
\end{eqnarray}
while Eq~\ref{sp22} becomes (defining $\epsilon_F = T_A$ as the position of the level in the
limit $N\rightarrow \infty$)
\begin{eqnarray}
T_A = E_0 - \frac{N\Gamma}{\pi} \left[\frac{\Gamma_L}{\Gamma}\ln\frac{|T_A - \mu_L|}{D}
+ \frac{\Gamma_R}{\Gamma}\ln\frac{|T_A - \mu_R|}{D}  \right] \label{sp22a}
\end{eqnarray}
Let us define
\begin{eqnarray}
m = \frac{N \Gamma}{\pi T_A} \label{mu}\\
{m}_V = m \left[\frac{{\Gamma}_L/\Gamma}{1 - \mu_L/T_A} + \frac{\Gamma_R/\Gamma}{1 - \mu_R/T_A} \right]
\label{muV}
\end{eqnarray}
$m$ should not to be confused with the label for the spin projection. 
Note that in equilibrium, $\mu_L = \mu_R =0$, ${m}_V = m$.  
In terms of these variables, Eq.~\ref{sp11a} implies the following for the saddle point solution for $b_{cl}$
\begin{eqnarray}
b_{cl}^2 = b_{sp}^2= \frac{\tilde{\Gamma}}{\Gamma} = \frac{1}{1 + {m}_V} \label{sp11b}
\end{eqnarray}
whereas the impurity charge density is
\begin{eqnarray}
n_F = -i \frac{N}{2}Tr\left[G^K_{mf}\right] + \frac{N}{2}
= 1- \frac{\tilde{\Gamma}}{\Gamma} = \frac{m_V}{1 + m_V} \label{sp11c}
\end{eqnarray}
Note that in the Kondo limit, $m_V \gg 1$ so that $n_F \rightarrow 1$.

\subsection{Solution for $T_A$}

We solve Eq.~\ref{sp22a} when $-E_0 \gg T_A$. 
Writing 
\begin{eqnarray}
T_A= T_A^0 + \delta T_A
\end{eqnarray}
where 
\begin{eqnarray}
T_A^0 = D e^{-\frac{\pi |E_0|}{N \Gamma}}\label{TA0}
\end{eqnarray}
is the equilibrium solution for the impurity level, Eq.~\ref{sp22a} becomes
\begin{eqnarray}
|T_A^0 + \delta T_A - \mu_L|^{\Gamma_L/\Gamma}
|T_A^0 + \delta T_A - \mu_R|^{\Gamma_R/\Gamma} = T_A^0
\label{solTA}
\end{eqnarray}
For small voltages, $|\delta T_A|, |\mu_{L,R}| << T_A^0$, a Taylor expansion leads to 
the following expression for the change in $T_A$ due to bias,
\begin{eqnarray}
\delta T_A = \left( \frac{\Gamma_L \mu_L}{\Gamma} + \frac{\Gamma_R \mu_R}{\Gamma}\right)
+ \frac{1}{2 T_A^0} \frac{\Gamma_L \Gamma_R}{\Gamma^2}\left( \mu_L - \mu_R\right)^2
\label{dTA1}
\end{eqnarray}

\subsection{Mean field impurity susceptibility}
We now turn to the evaluation of the voltage dependence of the impurity susceptibility.
The spin-response function at the mean-field level is given by 
\begin{equation}
\chi^R_{mf}(\Omega) = \frac{i}{2} \sum_{m}\left(g\mu_B m \right)^2 \int \frac{d\omega}{2\pi}
\left[G^R_{mf}(\omega + \Omega)G^K_{mf}(\omega) + G^K_{mf}(\omega + \Omega) G^A_{mf}(\omega)\right] \label{chimf}
\end{equation}
Using the identity
$\sum_{m=-J \ldots J} m^2 = \frac{2J + 1}{3} J(J+1) = \frac{N}{3} J(J+1)$, 
the spin susceptibility which is the zero frequency spin-response function becomes,
\begin{eqnarray}
\chi_{sp}= \chi^R_{mf}(\Omega=0) = \frac{g^2 \mu_B^2}{3} J(J+1) \left( \frac{N \tilde{\Gamma}}{\pi}\right)
\left[ \frac{\Gamma_L/\Gamma}{\left(T_A - \mu_L\right)^2 + \tilde{\Gamma}^2} + \frac{\Gamma_R/\Gamma}
{\left(T_A- \mu_R \right)^2 + \tilde{\Gamma}^2}\right]
\end{eqnarray}
For $N\rightarrow\infty$ we may drop terms of ${\cal O}(\tilde{\Gamma}^2)$, 
\begin{eqnarray}
&&\chi_{sp} = \frac{g^2 \mu_B^2}{3} J(J+1) \frac{m}{(1 + m_V)T_A } \sum_{i=L,R} 
\frac{\Gamma_i/\Gamma}{\left(1 - \mu_i/T_A\right)^2} \label{chisp} 
\end{eqnarray}
Taylor expanding Eq.~\ref{chisp} in powers of $\frac{\mu_{L,R}}{T_A^0}$ and defining
\begin{eqnarray}
 m_0 = \frac{N \Gamma }{\pi T_A^0} \label{m0def}
\end{eqnarray}
we find the following voltage dependence of the susceptibility at saddle-point, 
\begin{eqnarray}
\chi_{sp} = \frac{g^2 \mu_B^2}{3}J(J+1) \frac{m_0}{T_A^0(1 + m_0)}\left[ 1 + 
\left( \frac{4 + 3 m_0}{1 + m_0}\right)
\frac{\Gamma_L \Gamma_R}{2\Gamma^2}\left( \frac{\mu_L - \mu_R}{T_A^0}\right)^2 + \ldots \right]
\label{chisp1}
\end{eqnarray}
In the Kondo limit, $m_0 \gg 1$, or the equilibrium charge on the level 
$n_F=\frac{m_0}{1 + m_0}\rightarrow 1$.
In this case the static susceptibility becomes
\begin{eqnarray}
\chi_{sp}^{n_F=1} \rightarrow \frac{g^2 \mu_B^2}{3 T_A^0}J(J+1) \left[ 1 + 1.5
\frac{\Gamma_L \Gamma_R}{\Gamma^2}
\left( \frac{\mu_L - \mu_R}{T_A^0}\right)^2 + \ldots \right]
\label{chisp2}
\end{eqnarray}

\subsection{Mean-field conductance}

The current is given by~\cite{Wingreen94} 
\begin{eqnarray}
I = \frac{ie}{\hbar} 
\frac{2\Gamma_L \Gamma_R}{\Gamma}
\sum_m \int \frac{d\omega}{2\pi} \left(f(\omega-\mu_L) -f(\omega -\mu_R) \right) 
\left[G^R_{f_m,b}-G^A_{f_m,b}\right]\left(\omega\right) \label{I1} 
\end{eqnarray}
where 
\begin{eqnarray}
&&G^R_{f_m,b}\left(t,t^{\prime}\right) = -i 
T\langle b_-^{\dagger}(t) f_{m-}(t) f_{m-}^{\dagger}(t^{\prime}) b_-({t^{\prime}})\rangle 
- i \langle f_{m+}^{\dagger}(t^{\prime})b_+(t^{\prime}) b_-^{\dagger}(t) f_{m-}(t)\rangle
\label{spec}
\end{eqnarray}
Within mean-field, $b_{\pm}$ are constants in time and equal to the saddle point value given in
Eq.~\ref{sp11b}.
Thus at zero temperature Eq.~\ref{I1} becomes,
\begin{eqnarray}
I_{mf} = \frac{Ne}{h} \frac{4\Gamma_L \Gamma_R}{\Gamma}
\left(\frac{\tilde{\Gamma}}{\Gamma}\right) 
\left[\tan^{-1}\left(\frac{\mu_L-T_A}{\tilde{\Gamma}}\right) 
- \tan^{-1}\left(\frac{\mu_R-T_A}{\tilde{\Gamma}}\right) \right]
\end{eqnarray}
Let us set $\mu_L = eV/2, \mu_R = -eV/2$. 
The zero-bias conductance depends only on the equilibrium properties of the spectral density and is given by,
\begin{eqnarray}
&&G_{sp}\left(V=0\right) = \frac{\partial I_{mf}}{\partial V}|_{V=0} 
=\frac{N e^2}{h}\frac{4\Gamma_L \Gamma_R}{\Gamma^2} \frac{\tilde{\Gamma}^2}{T_A^2} \nonumber \\
&&= \frac{N e^2}{h}\frac{4\Gamma_L \Gamma_R}{\Gamma^2} \left(\frac{\pi n_F(V=0)}{N}\right)^2
\label{Gsp2}
\end{eqnarray}
The non-linearity in the conductance arises due to 
the frequency and voltage dependence of the spectral density (namely 
the voltage dependence of its position $T_A$ and its width $\tilde{\Gamma}$).
We find the following expression for the non-linear conductance,
\begin{eqnarray}
&&G_{sp}(V) = \frac{\partial I_{mf}}{\partial V} = \nonumber \\
&&G_{sp}(V=0)\left[1- 
\left(\frac{\Gamma_L -\Gamma_R}{\Gamma}\right)\left(\frac{2V}{T_A^0}\right)
-\frac{12\Gamma_L \Gamma_R}{\Gamma^2}\left(\frac{V}{T_A^0}\right)^2 +3
\left(\frac{V}{T_A^0}\right)^2 -\frac{3m_0}{1+m_0}\frac{\Gamma_L \Gamma_R}{\Gamma^2} 
\left(\frac{V}{T_A^0}\right)^2\right]\label{Gsp1a}
\end{eqnarray}
The above implies that
for asymmetric coupling to the leads ($\Gamma_L \neq \Gamma_R$), the
conductance shows a rectification type behavior, {\sl i.e.} $G_{sp}(V) \neq G_{sp}(-V)$. Whereas for symmetric
couplings to the leads, the conductance reduces to
\begin{eqnarray}
 G_{sp}(V;{\Gamma_L=\Gamma_R})= G_{sp}(V=0)
\left[1-\frac{3}{4}\frac{m_0}{1+m_0}
\left(\frac{V}{T_A^0}\right)^2\right] \label{Gsp1b}
\end{eqnarray}
The conductance in the Kondo limit can be obtained by setting $m_0 \gg 1$ in Eq.~\ref{Gsp1a},~\ref{Gsp1b}. Thus
for symmetric couplings, we get 
\begin{eqnarray}
 G_{sp}^{n_F=1}(V;\Gamma_L =\Gamma_R)= G_{sp}(V=0)
\left[1-\frac{3}{4}
\left(\frac{V}{T_A^0}\right)^2\right] \label{Gsp1c}
\end{eqnarray}

The main results of this section are the expressions for the static susceptibility 
(Eq.~\ref{chisp1},~\ref{chisp2}), 
and the conductance (Eqns.~\ref{Gsp1a},~\ref{Gsp1b},~\ref{Gsp1c}). 
In the rest of the paper we will study how these results are modified 
when fluctuations to ${\cal O}\left(\frac{1}{N}\right) $ are taken into account. 

\section{Fluctuations about mean-field} \label{fluc}

We now turn to the computation of how the saddle-point Eqns~\ref{sp11} and~\ref{sp22} 
get modified when 
fluctuations are included. Formally the steps involved are to write 
$b_{cl} \rightarrow b_{sp} + b_{cl}$, 
$b_{cl}^* \rightarrow b_{sp} + b_{cl}^*$, 
$b_q \rightarrow \bar{b}_q + b_q$, $b^*_q \rightarrow \bar{b}_q + b^*_q$. 
Then we integrate out all the fermionic and bosonic fields $b_{cl},b_{cl}^*,b_q,b_q^*$ 
obtaining a resulting action that depends only on $S_K = S_K(b_{sp}, \bar{b}_q,\lambda_{cl},
\lambda_q)$.
Each of the variables $x =b_{sp}, \bar{b}_q, \lambda_{cl},\lambda_q$ are then determined by
requiring that $\frac{\delta S_K(x,y\ldots)}{\delta x}=0$. Of course, the bosonic and 
fermionic fields cannot be
integrated out exactly. This is therefore done perturbatively in $\frac{1}{N}$. Moreover, 
as discussed in
Section~\ref{meanfield}, the saddle point equations with
respect to the classical fields $\frac{\delta S_K}{\delta b_{sp}}=0,
\frac{\delta S_K}{\delta \lambda_{cl}}=0$ are always satisfied if all the quantum fields 
$\lambda_q=\bar{b}_q=0$. Thus to obtain the quantum saddle-points, it suffices to expand $S_K$ to
only the leading power in the quantum fields $\lambda_q,\bar{b}_q$. 
To make the computation simple, we will carry this out
separately for the saddle-point equation for $\lambda$ and $b_{sp}$.

\begin{figure}
\includegraphics[totalheight=5cm,width=5cm]{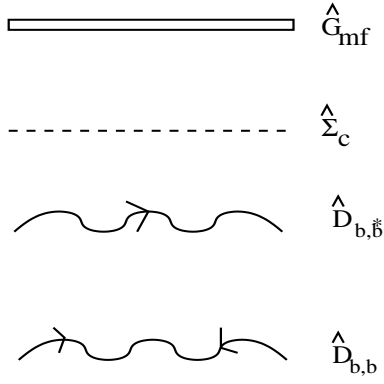}
\caption{
Diagrams representing the mean-field Green's function $\hat{G}_{mf}$, the self-energy due to coupling to 
leads $\hat{\Sigma}_c$, and the
bosonic propagator $\hat{D}$. Each has a $2 \times 2$ Keldysh structure. 
}
\label{fig1}
\end{figure}

\subsection{Saddle point equation for $\lambda$} 
\begin{figure}
\includegraphics[totalheight=4cm,width=4cm]{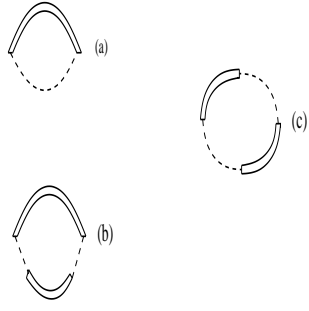}
\caption{The bosonic self-energies corresponding to (a). $\Pi$; (b). $\delta \Pi^{(1)}$ and (c). $\delta \Pi^{(2)}$ 
in text}
\label{fig2}
\end{figure}

In order to compute $1/N$ corrections to the saddle point equation for $\lambda$ (Eq.~\ref{sp11}), 
we write $b_{cl} \rightarrow b_{sp} + b_{cl}$, and expand the action
in Eq.~\ref{SK3} in powers of $b_{cl},b_q$ and $\lambda_q$.
To achieve this we first integrate out the fermion fields in Eq.~\ref{SK3} to obtain
\begin{eqnarray}
&&S_K = -i N Tr\ln
\left[G_{mf}^{-1} - \lambda_q \tau_x - b_{sp} \Sigma_c \left(b^*_{cl}\tau_0 + b^*_q \tau_x \right)
-\left(b_{cl}\tau_0 + b_q \tau_x \right)\Sigma_c b_{sp} -
\left(b_{cl}\tau_0 + b_q \tau_x \right) \Sigma_c \left(b^*_{cl}\tau_0 + b^*_q \tau_x \right) \right]
\nonumber \\
&& + 2\begin{pmatrix} {b}^*_{cl} & {b}^*_{q} \end{pmatrix}
\begin{pmatrix}  -\lambda_q
&i\partial_t - \lambda_{cl}\\i \partial_t -\lambda_{cl} &  -\lambda_q \end{pmatrix}
\begin{pmatrix} b_{cl}\\ b_{q}\end{pmatrix} + 2 \lambda_q\left(1 - \frac{N}{2} \right)
- 2 \lambda_q b_{sp}^2 - 2 \lambda_qb_{sp} (b_{cl}+b_{cl}^*)  - 2 \lambda_{cl}b_{sp} \left(b_q +b_q^*\right)
\label{SK4}
\end{eqnarray}
Let us define,
\begin{eqnarray}
\tilde{G}^{-1}_{mf}(\lambda_q) = G^{-1}_{mf} - \lambda_q \tau_x
\end{eqnarray}
The solution to the above equation to leading order in $\lambda_q$ is
\begin{eqnarray}
\tilde{G}_{mf} = G_{mf} + \lambda_q  G_{mf} \tau_x G_{mf} = G_{mf} + \lambda_q  \delta G_{mf}
\label{dgmf}
\end{eqnarray}
where we define
\begin{eqnarray}
\delta G_{mf} = G_{mf} \tau_x G_{mf}
\label{dgmf2}
\end{eqnarray}
Expanding Eq.~\ref{SK4} to quadratic order in the fluctuating fields $b_{q,cl}$ we get
\begin{eqnarray}
&&S_K = -i N Tr\ln\tilde{G}^{-1}_{mf}(\lambda_q)
+i N b_{sp}Tr\left[\tilde{G}_{mf}(\lambda_q)\left(\Sigma_c \left(b^*_{cl}\tau_0 + b^*_q \tau_x \right)
+\left(b_{cl}\tau_0 + b_q \tau_x \right)\Sigma_c \right)\right] \nonumber \\
&& + \frac{i N b_{sp}^2}{2}Tr\left[\left(\tilde{G}_{mf}(\lambda_q)\left(\Sigma_c
\left(b^*_{cl}\tau_0 + b^*_q \tau_x \right)
+\left(b_{cl}\tau_0 + b_q \tau_x \right)\Sigma_c \right) \right)^2 \right]\nonumber \\
&&+i N
Tr\left[\tilde{G}_{mf}(\lambda_q)\left(b_{cl}\tau_0 + b_q \tau_x \right) \Sigma_c \left(b^*_{cl}\tau_0 + b^*_q \tau_x \right)
\right]+ 2\begin{pmatrix} {b}^*_{cl} & {b}^*_{q} \end{pmatrix}
\begin{pmatrix}  -\lambda_q
&i\partial_t - \lambda_{cl}\\i \partial_t -\lambda_{cl} &  -\lambda_q \end{pmatrix}
\begin{pmatrix} b_{cl}\\ b_{q}\end{pmatrix} \nonumber \\
&&+ 2 \lambda_q\left(1 - \frac{N}{2} \right)
- 2 \lambda_q b_{sp}^2 - 2 \lambda_q  b_{sp}\left(b_{cl}+b^*_{cl}\right) - 2 \lambda_{cl}b_{sp}
\left(b_q + b_q^*\right)
\label{SK5}
\end{eqnarray}
Collecting all terms upto quadratic order in the bosonic fields, we rewrite the action as below,
\begin{eqnarray}
&&S_K = -i N Tr\ln\tilde{G}^{-1}_{mf}(\lambda_q)
+i N b_{sp}Tr\left[\tilde{G}_{mf}(\lambda_q)\left(\Sigma_c \left(b^*_{cl}\tau_0 + b^*_q \tau_x \right)
+\left(b_{cl}\tau_0 + b_q \tau_x \right)\Sigma_c \right)\right] \nonumber \\
&&+ 2\begin{pmatrix} {b}^*_{cl} & {b}^*_{q} \end{pmatrix}
\left[\begin{pmatrix}  -\lambda_q
&i\partial_t - \lambda_{cl}
\\i \partial_t -\lambda_{cl} &  -\lambda_q \end{pmatrix} - \Pi - \delta\Pi^{(1)} -
\lambda_q\left( \delta \Pi_q + \delta \Pi_q^{(1)} \right)\right]
\begin{pmatrix} b_{cl}\\ b_{q}\end{pmatrix} \nonumber \\
&&+ 2\begin{pmatrix} {b}^*_{cl} & {b}^*_{q} \end{pmatrix}
\left[- \delta\Pi^{(2)} -
\lambda_q\delta \Pi_q^{(2)} \right]
\begin{pmatrix} b^*_{cl}\\ b^*_{q}\end{pmatrix} +
2\begin{pmatrix} {b}_{cl} & {b}_{q} \end{pmatrix}
\left[- \delta\Pi^{(2)} -
\lambda_q\delta \Pi_q^{(2)} \right]
\begin{pmatrix} b_{cl}\\ b_{q}\end{pmatrix} \nonumber \\
&&+ 2 \lambda_q\left(1 - \frac{N}{2} \right)
- 2 \lambda_q b_{sp}^2 - 2 \lambda_q  b_{sp}\left(b_{cl}+b^*_{cl}\right) - 2 \lambda_{cl}b_{sp}
\left(b_q + b_q^*\right)
\label{SK5b}
\end{eqnarray}
The above shows that the bosons
due their interaction with fermions acquire the self-energies $\Pi=\begin{pmatrix} 0 & \Pi^A \\\Pi^R & \Pi^K 
\end{pmatrix}
$ , $\delta\Pi^{(1,2)}=\begin{pmatrix} 0 & \delta \Pi^{A(1,2)} \\ \delta \Pi^{R(1,2)} & \delta \Pi^{K(1,2)} \end{pmatrix}$. The diagrams corresponding to $\Pi,\delta \Pi^{(1,2)}$ are shown in Fig~\ref{fig2} (where the propagators 
are defined in Fig~\ref{fig1}). The bosonic self-energy $\Pi$ is,
\begin{eqnarray}
&&\Pi(t,t^{\prime}) =
\frac{-iN}{2} \begin{pmatrix}Tr^{\prime}\left[\tau_0 G_{mf}(t,t^{\prime})\tau_0 \Sigma_c(t^{\prime},t)\right]   & Tr^{\prime}\left[\tau_0 G_{mf}(t,t^{\prime})\tau_x \Sigma_c(t^{\prime},t)\right]
\\ Tr^{\prime}\left[\tau_x G_{mf}(t,t^{\prime})\tau_0 \Sigma_c(t^{\prime},t)\right]  &
Tr^{\prime}\left[\tau_x G_{mf}(t,t^{\prime})\tau_x \Sigma_c(t^{\prime},t) \right]  \end{pmatrix} \label{Pidef}
\end{eqnarray}
where $Tr^{\prime}$ implies trace over only the Keldysh indices. Note that from causality the upper-left term in 
Eq.~\ref{Pidef} is zero. 
Explicit expressions for $\Pi$ are given in Appendix~\ref{Pieval}.
The other self-energies are,
\begin{eqnarray}
&&\delta \Pi^{A(1)}(t,t^{\prime}) = \frac{-iNb_{sp}^2}{4} \label{deltaPi1A}\\
&&\left[\left(G^K_{mf}\right)_{t,t^{\prime}}\left(\Sigma^R_cG^R_{mf}\Sigma^R_c\right)_{t^{\prime},t}
+\left(G^A_{mf}\right)_{t,t^{\prime}} \left(\Sigma^R_cG^R_{mf}\Sigma^K_c\right)_{t^{\prime},t}
+\left(G^A_{mf}\right)_{t,t^{\prime}} \left(\Sigma^R_cG^K_{mf}\Sigma^A_c\right)_{t^{\prime},t}
+ \left(G^A_{mf}\right)_{t,t^{\prime}}\left(\Sigma^K_cG^A_{mf}\Sigma^A_c\right)_{t^{\prime},t}\right]
\nonumber \\
&&\delta \Pi^{R(1)}(t,t^{\prime}) = \frac{-iNb_{sp}^2}{4} \label{deltaPi1R}\\
&&\left[\left(G^R_{mf}\right)_{t,t^{\prime}} \left(\Sigma^R_cG^R_{mf}\Sigma^K_c\right)_{t^{\prime},t}
+\left(G^R_{mf}\right)_{t,t^{\prime}} \left(\Sigma^R_cG^K_{mf}\Sigma^A_c\right)_{t^{\prime},t}
+ \left(G^R_{mf}\right)_{t,t^{\prime}}\left(\Sigma^K_cG^A_{mf}\Sigma^A_c\right)_{t^{\prime},t}
+ \left(G^K_{mf}\right)_{t,t^{\prime}}\left(\Sigma^A_cG^A_{mf}\Sigma^A_c\right)_{t^{\prime},t}\right]
\nonumber \\
&&\delta \Pi^{K(1)}(t,t^{\prime}) = \frac{-iNb_{sp}^2}{4} \label{deltaPi1K}\\
&&\{\left(G^K_{mf}\right)_{t,t^{\prime}}\left(\Sigma^R_cG^R_{mf}\Sigma^K_c\right)_{t^{\prime},t}
+\left(G^K_{mf}\right)_{t,t^{\prime}} \left(\Sigma^R_cG^K_{mf}\Sigma^A_c\right)_{t^{\prime},t}
+ \left(G^K_{mf}\right)_{t,t^{\prime}}\left(\Sigma^K_cG^A_{mf}\Sigma^A_c\right)_{t^{\prime},t}
+ \left(G^A_{mf}\right)_{t,t^{\prime}}\left(\Sigma^R_cG^R_{mf}\Sigma^R_c\right)_{t^{\prime},t}
\nonumber \\
&& + \left(G^R_{mf}\right)_{t,t^{\prime}}\left(\Sigma^A_cG^A_{mf}\Sigma^A_c\right)_{t^{\prime},t}
\}
\nonumber
\end{eqnarray}
which are of ${\cal O}(1/N^2)$ and therefore will be dropped. The anomalous boson self-energies are the following
\begin{eqnarray}
&&\delta \Pi^{R(2)} = \frac{-iNb_{sp}^2}{4} \label{PiRbb}\\
&&\left[\left(G^R_{mf}\Sigma^R_c\right)_{t,t^{\prime}}\left(G^R_{mf}\Sigma^K_c\right)_{t^{\prime}t}
+\left(G^R_{mf}\Sigma^R_c\right)_{t,t^{\prime}}\left(G^K_{mf}\Sigma^A_c\right)_{t^{\prime}t}
+ \left(G^R_{mf}\Sigma^K_c\right)_{t,t^{\prime}}\left(G^A_{mf}\Sigma^A_c\right)_{t^{\prime}t}
+ \left(G^K_{mf}\Sigma^A_c\right)_{t,t^{\prime}}\left(G^A_{mf}\Sigma^A_c\right)_{t^{\prime}t}\right]
\nonumber \\
&&\delta \Pi^{A(2)} = \frac{-iNb_{sp}^2}{4} \label{PiAbb}\\
&&\left[\left(G^A_{mf}\Sigma^A_c\right)_{t,t^{\prime}}\left(G^R_{mf}\Sigma^K_c\right)_{t^{\prime}t}
+\left(G^R_{mf}\Sigma^K_c\right)_{t,t^{\prime}}\left(G^R_{mf}\Sigma^R_c\right)_{t^{\prime}t}
+ \left(G^A_{mf}\Sigma^A_c\right)_{t,t^{\prime}}\left(G^K_{mf}\Sigma^A_c\right)_{t^{\prime}t}
+ \left(G^K_{mf}\Sigma^A_c\right)_{t,t^{\prime}}\left(G^R_{mf}\Sigma^R_c\right)_{t^{\prime}t}\right]
\nonumber \\
&&\delta \Pi^{K(2)} = \frac{-iNb_{sp}^2}{4} \label{PiKbb}\\
&&\{\left(G^R_{mf}\Sigma^K_c\right)_{t,t^{\prime}}\left(G^R_{mf}\Sigma^K_c\right)_{t^{\prime}t}
+\left(G^R_{mf}\Sigma^K_c\right)_{t,t^{\prime}}\left(G^K_{mf}\Sigma^A_c\right)_{t^{\prime}t}
+ \left(G^K_{mf}\Sigma^A_c\right)_{t,t^{\prime}}\left(G^R_{mf}\Sigma^K_c\right)_{t^{\prime}t}
+ \left(G^K_{mf}\Sigma^A_c\right)_{t,t^{\prime}}\left(G^K_{mf}\Sigma^A_c\right)_{t^{\prime}t}
\nonumber \\
&& + \left(G^A_{mf}\Sigma^A_c\right)_{t,t^{\prime}}\left(G^R_{mf}\Sigma^R_c\right)_{t^{\prime}t}
+ \left(G^R_{mf}\Sigma^R_c\right)_{t,t^{\prime}}\left(G^A_{mf}\Sigma^A_c\right)_{t^{\prime}t}
\}
\nonumber
\end{eqnarray}
and are at least of ${\cal O}(1/N)$. These will therefore not play a role in the ${\cal O}\left(\frac{1}{N}\right)$
corrections to the saddle point equations, but will be important later, when we evaluate the conductance.
The self-energies $\delta \Pi_q^{(2)}$ is also of ${\cal O}(1/N)$ and will be dropped from
further consideration.

Other self-energies needed for computing corrections to the saddle point equations are $\delta\Pi_q$ and 
$\delta\Pi_q^{1}$. We find,
\begin{eqnarray}
&&\delta \Pi_q(t,t^{\prime}) =  \begin{pmatrix} \delta\Pi^{z}_q & \delta \Pi^{A}_q \\ \delta \Pi^{R}_q &
\delta\Pi^{K}_q \end{pmatrix}
\\
&&=\frac{-iN}{2}
\begin{pmatrix}Tr^{\prime}\left[\tau_0 \delta G_{mf}(t,t^{\prime})\tau_0 \Sigma_c(t^{\prime},t)\right]
& Tr^{\prime}\left[\tau_0 \delta G_{mf}(t,t^{\prime})\tau_x \Sigma_c(t^{\prime},t)\right]
\\ Tr^{\prime}\left[\tau_x \delta G_{mf}(t,t^{\prime})\tau_0 \Sigma_c(t^{\prime},t)\right]  &
Tr^{\prime}\left[\tau_x \delta G_{mf}(t,t^{\prime})\tau_x \Sigma_c(t^{\prime},t) \right]  \end{pmatrix} \label{dPidef}
\end{eqnarray}
with $\delta G_{mf}$ defined in Eq.~\ref{dgmf2}.
Whereas, $\delta \Pi_q^{(1)}$ is (retaining
terms upto ${\cal O}(1/N)$),
\begin{eqnarray}
&&\delta \Pi^{(1)}_q(t,t^{\prime}) =  \begin{pmatrix} \delta\Pi^{z(1)}_q={\cal O}(1/N^2) & \delta \Pi^{A(1)}_q
\\ \delta \Pi^{R(1)}_q &
\delta\Pi^{K(1)}_q={\cal O}(1/N^2) \end{pmatrix}
\end{eqnarray}
where
\begin{eqnarray}
\delta \Pi_q^{R(1)} =\frac{-iN b_{sp}^2}{2}\left[
G^R_{mf}(t,t^{\prime}) \left(\Sigma^R_c \delta G^K_{mf} \Sigma^A_c \right)_{t^{\prime},t}
+ G^R_{mf}(t,t^{\prime}) \left(\Sigma^K_c \delta G^z_{mf} \Sigma^K_c \right)_{t^{\prime},t}
+  G^A_{mf}(t,t^{\prime}) \left(\Sigma^A_c \delta G^z_{mf} \Sigma^R_c \right)_{t^{\prime},t}\right] \\
\delta \Pi_q^{A(1)} =\frac{-iN b_{sp}^2}{2}\left[
G^A_{mf}(t,t^{\prime}) \left(\Sigma^R_c \delta G^K_{mf} \Sigma^A_c \right)_{t^{\prime},t}
+ G^A_{mf}(t,t^{\prime}) \left(\Sigma^K_c \delta G^z_{mf} \Sigma^K_c \right)_{t^{\prime},t}
+  G^R_{mf}(t,t^{\prime}) \left(\Sigma^A_c \delta G^z_{mf} \Sigma^R_c \right)_{t^{\prime},t}\right]
\end{eqnarray}

We now integrate out the bosonic fields in the action Eq.~\ref{SK5b}. The ${\cal O} (\lambda_q^0)$ 
term cancels the last
term because of the saddle point
condition Eq~\ref{sp2}, whereas the ${\cal O}(\lambda_q^1)$ term is also first order in the fluctuating
bosonic fields $b_{q,cl}$.
Thus on integrating out $b_{q,cl}$, this term gives a term in
the Keldysh action which is $\lambda_q^2$ and therefore does not affect the classical saddle point
solutions. Following these steps we obtain,
\begin{eqnarray}
&&S_K = -i N Tr\ln\tilde{G}^{-1}_{mf}(\lambda_q)
+i Tr\ln\left[D_0^{-1} - \Pi - \lambda_q \tau_0 -\lambda_q \left(\delta\Pi_q + \delta \Pi^{(1)}_q\right)
\right] \nonumber\\
&&+ 2 \lambda_q\left(1 - \frac{N}{2} \right)
- 2 \lambda_q b_{sp}^2 + {\cal O}(\lambda_q^2) + {\cal O}(1/N^2)
\label{SK6}
\end{eqnarray}
where i$\langle b_{a} b^*_{b}\rangle = \frac{1}{2}D_0$ is the bare bosonic propagator.

Now we may differentiate Eq~\ref{SK6} with respect to $\lambda_q$ and set all quantum fields to
zero in the resultant expression to obtain,
\begin{eqnarray}
2\left(1-\frac{N}{2}\right)  - 2 b_{sp}^2 + i N Tr\left[G^K_{mf}\right]
-i Tr\left[ D_{b,b^*} + D_{b,b^*}\left(\delta \Pi_q + \delta \Pi_q^{(1)}\right)\right] = 0
\label{splamN1}
\end{eqnarray}
Upto ${\cal O}(1/N)$ only a subset of terms in Eq.~\ref{splamN1} need to be kept. Collecting these,
\begin{eqnarray}
&&2\left(1-\frac{N}{2}\right)  - 2 b_{sp}^2 + i N Tr\left[G^K_{mf}\right] - i Tr D^K_{b,b^*}
=
\frac{N}{2} \int \frac{d\epsilon}{2\pi} \int \frac{d\omega}{2\pi} \label{splamN2}\\
&&\left[D^A_{b,b^*}(\epsilon) G_{mf}^A(\epsilon+\omega)
G_{mf}^K(\epsilon+\omega)+ D^R_{b,b^*}(\epsilon) G_{mf}^K(\epsilon+\omega)
G_{mf}^R(\epsilon+\omega)+ D^K_{b,b^*}(\epsilon)G_{mf}^A(\epsilon+\omega)
G_{mf}^R(\epsilon+\omega) \right]\Sigma^K(\omega)\nonumber
\end{eqnarray}

Substituting for the fermionic and bosonic Green's functions (Eq.~\ref{GRmfdef},~\ref{GKmfdef},~\ref{eFdef},~\ref{DKdef1},~\ref{ReDR},~\ref{ImDR}), we get
\begin{eqnarray}
&&1 - b_{sp}^2 - \frac{N\tilde{\Gamma}}{\pi \epsilon_F}\sum_{a=L,R} \frac{\Gamma_a/\Gamma}
{1-\mu_a/\epsilon_F}  = \label{Ncorrlam}\\
&&\frac{N \Gamma \tilde{\Gamma}}{\pi^2 T_A^2}
\sum_{a,b=L,R}
\frac{\Gamma_a \Gamma_b}{\Gamma^2}\int_{-(\mu_a -\mu_b)/T_A}^{D/T_A} dx 
\frac{\left(\frac{1}{1-\frac{\mu_a}{T_A}}
-\frac{1}{1+x -\frac{\mu_b}{T_A}} \right)}
{\left(x +m\sum_{i=L,R}\frac{\Gamma_i}{\Gamma} 
\ln{|1+\frac{x}{1 -\mu_i/T_A}|}\right)^2} 
\left(1 +m \sum_{i} \frac{\Gamma_i}{\Gamma}\frac{1}{1+ x - \mu_i/T_A }
\right)\nonumber \\
&&+\frac{N \Gamma \tilde{\Gamma}}{\pi^2T_A^2}
\sum_{a,b=L,R}
\frac{\Gamma_a \Gamma_b}{\Gamma^2}\int_{-(\mu_a -\mu_b)/T_A}^{D/T_A} d x 
\frac{\left(\frac{1}{(1-\frac{\mu_a}{T_A})^2}
-\frac{1}{(1 +x -\frac{\mu_b}{T_A})^2} \right)}
{x + m \sum_{i}\frac{\Gamma_i}{\Gamma} 
\ln{|1+ \frac{x}{1-\mu_i/T_A}|}} 
\nonumber
\end{eqnarray}

We use the identity
\begin{eqnarray}
&&1 + m \sum_i \frac{\Gamma_i/\Gamma}{1 + x -\mu_i/T_A} = \sum_i \frac{\Gamma_i}{\Gamma}
\left[ 1 + \frac{m}{1-\mu_i/T_A}- \frac{m}{1-\frac{\mu_i}{T_A}}
\left(\frac{x}{1+x-\mu_i/T_A}\right)\right] \nonumber \\
&& = 1 + m_V - m x \sum_i 
\left(\frac{\Gamma_i/\Gamma}{(1+x-\mu_i/T_A)(1-\frac{\mu_i}{T_A})}\right)
\end{eqnarray}
and for convenience introduce the following short-hand,
\begin{eqnarray}
L_p^{a,b} = \frac{\Gamma_a \Gamma_b}{\Gamma^2} \int_{-(\mu_a -\mu_b)/T_A}^{D/T_A} d x 
\frac{\left(\frac{1}{(1-\frac{\mu_a}{T_A})^p}
-\frac{1}{(1 +x -\frac{\mu_b}{T_A})^p} \right)}
{x + m \sum_{i}\frac{\Gamma_i}{\Gamma} 
\ln{|1+ \frac{x}{1-\mu_i/T_A}|}} 
\label{Ldef}
\end{eqnarray}
\begin{eqnarray}
&&k^{a,b} = \frac{\Gamma_a \Gamma_b}{\Gamma^2} \int_{-(\mu_a -\mu_b)/T_A}^{D/T_A} d x 
\frac{\left(\frac{x +  (\mu_a -\mu_b)/T_A}{(1-\mu_a/T_A)(1+x -\mu_b/T_A)}\right) 
x \sum_i \frac{\Gamma_i/\Gamma}{(1 + x - \mu_i/T_A)(1-\mu_i/T_A)}  }
{(x + m \sum_{i}\frac{\Gamma_i}{\Gamma} 
\ln{|1+ \frac{x}{1-\mu_i/T_A}|})^2} \label{kdef} \\
&&I^{a,b} = \frac{\Gamma_a \Gamma_b}{\Gamma^2} \int_{-(\mu_a -\mu_b)/T_A}^{D/T_A} d x 
\frac{\left(\frac{x +  (\mu_a -\mu_b)/T_A}{(1-\mu_a/T_A)(1+x -\mu_b/T_A)}\right) 
}{(x + m \sum_{i}\frac{\Gamma_i}{\Gamma} 
\ln{|1+ \frac{x}{1-\mu_i/T_A}|})^2} 
\label{Idef}
\end{eqnarray}
Note that the functions $I^{a,b}$ are infrared divergent. In equilibrium ($\mu_L =\mu_R = 0$) these have
a logarithmic divergence, while out of equilibrium the $I^{a,b}$ have a more severe $1/x$ divergence. 
As shown in equilibrium by Read et al~\cite{Read85,Read87},
in the computation of physical observables the $I^{a,b}$ appear in such a way
as to exactly cancel the divergences. For the out-of-equilibrium calculation as well we find an
exact cancellation of divergences, so that all physical observables are well defined. 

In terms of the above symbols, Eq.~\ref{Ncorrlam} becomes
\begin{eqnarray}
1 - b_{sp}^2 - \frac{N\tilde{\Gamma}}{\pi \epsilon_F}\sum_{a=L,R} \frac{\Gamma_a/\Gamma}
{1-\mu_a/\epsilon_F}  =
\frac{N \Gamma \tilde{\Gamma}}{\pi^2 T_A^2}
\sum_{a,b=L,R} \left[-m k^{a,b} + (1 + m_V) I^{a,b} +  L_2^{a,b} \right] \label{Ncorrlam2}
\end{eqnarray}

The l.h.s of the above equation can be further arranged as follows by writing $\epsilon_F = T_A + \beta/N$
\begin{eqnarray}
\frac{N\tilde{\Gamma}}{\pi \epsilon_F} \sum_{a}\frac{\Gamma_a/\Gamma}{1 - \mu_a/\epsilon_F}
= 
\frac{N\tilde{\Gamma}}{\pi T_A} \sum_{a}\frac{\Gamma_a/\Gamma}{1 - \mu_a/T_A} -\frac{\beta}{N T_A}
\frac{N\tilde{\Gamma}}{\pi T_A} \sum_{a}\frac{\Gamma_a/\Gamma}{(1 - \mu_a/T_A)^2}
\label{epsiloncorr}
\end{eqnarray}
The $1/N$ correction to $\epsilon_F$ is carried out in the next subsection. Using the result
for $\beta$ derived there (Eq.~\ref{betadef2}), the above equation is rewritten as
\begin{eqnarray}
\frac{N\tilde{\Gamma}}{\pi \epsilon_F} \sum_{a}\frac{\Gamma_a/\Gamma}{1 - \mu_a/\epsilon_F}
= \frac{N\tilde{\Gamma}}{\pi T_A} \sum_{a}\frac{\Gamma_a/\Gamma}{1 - \mu_a/T_A} -
\left(\frac{m^2}{N(1+m_V)}
\sum_{a,b} L_1^{a,b}\right)\frac{N\tilde{\Gamma}}{\pi T_A} 
\sum_{a}\frac{\Gamma_a/\Gamma}{(1 - \mu_a/T_A)^2}
\label{test1}
\end{eqnarray}

Substituting Eq.~\ref{test1} into Eq.~\ref{Ncorrlam2} we obtain the 
following expression for $\tilde{\Gamma}$ upto ${\cal O}(1/N)$
\begin{eqnarray}
&&\frac{\tilde{\Gamma}}{\Gamma} =  \label{NcorrGam1}\\
&&1 - \frac{m_V}{1 + m_V}
\left[1 + \frac{1}{N}\sum_{a,b}\left(\frac{m^2/m_V}{1 + m_V}L_2^{a,b} 
- \frac{m^3/m_V}{1 + m_V}k^{a,b} - \frac{m^3/m_V}{(1 + m_V)^2}L_1^{a,b}
\left(\sum_{i}\frac{\Gamma_i/\Gamma}{(1 - \mu_i/T_A)^2}\right) \right) \right] 
- \frac{1}{N} \frac{m^2}{1 + m_V} \sum_{a,b}I^{a,b} \nonumber 
\end{eqnarray}

It is convenient to introduce the following simplified notation
\begin{eqnarray}
k = \sum_{a,b}k^{a,b}\\
I = \sum_{a,b}I^{a,b} \\
L_p = \sum_{a,b}L^{a,b}_p \label{Lpdef}\\
S_{p=1,2,3} = \sum_i \frac{\Gamma_i/\Gamma}{(1 - \mu_i/T_A)^p}\label{Sp}
\end{eqnarray}
Then,

\begin{eqnarray}
\frac{\tilde{\Gamma}}{\Gamma} =  
1 - \frac{m_V}{1 + m_V}
\left[1 + \frac{1}{N}\left(\frac{m^2/m_V}{1 + m_V}L_2 
- \frac{m^3/m_V}{1 + m_V}k - \frac{m^3/m_V}{(1 + m_V)^2}L_1 S_2\right) \right] 
- \frac{1}{N} \frac{m^2}{1 + m_V} I \label{NcorrGam2}
\end{eqnarray}
Note that in equilibrium when $\mu_L = \mu_R = 0$, Eq.~\ref{Lpdef} becomes
\begin{eqnarray}
L_p^{eq} = \int_{0}^{D/T_A} d x 
\frac{\left(1-\frac{1}{(1 +x)^p} \right)}{x + m  \ln{\left(1+x\right)}} 
\label{Lpeq}
\end{eqnarray}
which is an expression that will appear later in the voltage expansion for physical observables.

Thus the main result of this sub-section is Eq.~\ref{NcorrGam2} which is the $1/N$ correction to the
level broadening. 

\subsection{Saddle point equation for $b$}
In order to derive the $1/N$ corrections to the saddle point Eq.~\ref{spR}, we set
$\lambda_q=0$ in Eq.~\ref{SK3}, write $b_{cl} = b_{sp} + b_{cl},b^*_{cl} = b_{sp} + b^*_{cl} $,
$b_q = \bar{b}_q + b_q,b^*_q = \bar{b}_q + b^*_q $,
and expand to quadratic order in the fluctuating fields $b_{q,cl}$. Following this as before, we
integrate out the fermions and the bosons and obtain an action $S_K(\bar{b}_q)$. To obtain
the classical saddle point, we need $\frac{\delta S_K(\bar{b}_q)}{\delta \bar{b}_{q}}|_{\bar{b}_q=0}$.
We will now follow the above steps.

First we integrate out the electrons to obtain,
\begin{eqnarray}
&&S_K = -i N Tr\ln
\left[ G_{mf}^{-1} - b_{sp}\tau_0 \Sigma_c \bar{b}_q \tau_x - {\bar b}_q \tau_x \Sigma_c b_{sp}\tau_0
- {\cal O}(\bar{b}_q^2)
- \left(b_{sp}\tau_0 + \bar{b}_q\tau_x\right) \Sigma_c \left(b^*_{cl}\tau_0 + b^*_q \tau_x \right) \nonumber 
\right.\\
&&\left.-\left(b_{cl}\tau_0 + b_q \tau_x \right) \Sigma_c \left(b_{sp} \tau_0 + \bar{b}_q \tau_x \right)
-\left(b_{cl}\tau_0 + b_q \tau_x \right) \Sigma_c \left(b^*_{cl}\tau_0 + b^*_q \tau_x \right) \right]
\nonumber \\
&& + 2\begin{pmatrix} {b}^*_{cl} & {b}^*_{q} \end{pmatrix}
\begin{pmatrix} 0
&i \partial_t  - \lambda_{cl}\\i\partial_t -\lambda_{cl} & 0 \end{pmatrix}
\begin{pmatrix} b_{cl}\\ b_{q}\end{pmatrix}
- 4 \lambda_{cl}b_{sp}\bar{b}_q -  2 \lambda_{cl}b_{sp} \left(b_q+b_q^*\right)
- 2 \lambda_{cl}\bar{b}_q \left(b_{cl}+b_{cl}^*\right)\label{SK5a}
\end{eqnarray}
Expanding the above to quadratic order in the fluctuating fields we get
\begin{eqnarray}
&&S_K = i N b_{sp}\bar{b}_q Tr\left[ G_{mf}\{\tau_x \Sigma_c +
\Sigma_c \tau_x\}\right] + i N b_{sp}^2 \bar{b}_q\left(b_q^*
Tr\left[G^R_{mf}\Sigma^K_c + G^K_{mf}\Sigma^A_c \right] + b_q \left[
G^K_{mf}\Sigma^R_c +G^A_{mf}\Sigma^K_c\right] \right)\nonumber \\
&&+i N Tr\left[\left(b_{sp}b_q^*+ b_{cl}\bar{b}_q\right)\left(
G^R_{mf}\Sigma^K_c + G^K_{mf}\Sigma^A_c\right) + \left(b_q b_{sp}
+\bar{b}_q b_{cl}^*\right) \left(G^K_{mf}\Sigma^R_c +
G^A_{mf}\Sigma^K_c \right) \right]
\nonumber \\
&&+ i N\bar{b}_q \left(b_q + b_q^*\right) Tr\left[G^R_{mf}
\Sigma^A_c + G^A_{mf} \Sigma^R_c + G^K_{mf} \Sigma^K_c\right] +
2\begin{pmatrix} {b}^*_{cl} & {b}^*_{q} \end{pmatrix} \left[D_0^{-1}
- \Pi - \delta\Pi^{(1)} - b_{sp} \bar{b}_q \left(\delta
\Pi^{\prime}_q + \delta \Pi^{\prime(1)}_q \right) \right]
\begin{pmatrix} b_{cl}\\ b_{q}\end{pmatrix} \nonumber \\
&& + 2\begin{pmatrix} {b}^*_{cl} & {b}^*_{q} \end{pmatrix}
\left[- \delta\Pi^{(2)} -
b_{sp} \bar{b}_q \delta \Pi_q^{\prime (2)} \right]
\begin{pmatrix} b^*_{cl}\\ b^*_{q}\end{pmatrix} +
2\begin{pmatrix} {b}_{cl} & {b}_{q} \end{pmatrix}
\left[- \delta\Pi^{(2)} -
b_{sp} \bar{b}_q \delta \Pi_q^{\prime (2)} \right]
\begin{pmatrix} b_{cl}\\ b_{q}\end{pmatrix}\nonumber  \nonumber \\
&&- 4 \lambda_{cl}b_{sp}\bar{b}_q
-  2 \lambda_{cl}b_{sp} \left(b_q + b_q^*\right) - 2
\lambda_{cl}\bar{b}_q\left(b_{cl} + b_{cl}^*\right)  \label{SK7a}
\end{eqnarray}
with $\Pi$ defined in Eq.~\ref{Pidef}, and the components of $\delta\Pi^{1}$ defined in 
Eq.~\ref{deltaPi1A},~\ref{deltaPi1R},~\ref{deltaPi1K}, and those of $\delta \Pi^2$ defined in 
Eq.~\ref{PiRbb},~\ref{PiAbb} and~\ref{PiKbb}. Moreover,  
the $\delta \Pi^{\prime}_q$ are given by,
\begin{eqnarray}
\delta \Pi^{\prime}_q(t,t^{\prime}) =
\frac{-iN}{2}\begin{pmatrix}Tr\left[\left(G_{mf}(\Sigma_c \tau_x +
\tau_x \Sigma_c)G_{mf}\right)(t,t^{\prime})\Sigma_c(t^{\prime},t)
\right] & Tr\left[\left(G_{mf}(\Sigma_c \tau_x + \tau_x
\Sigma_c)G_{mf}\right)(t,t^{\prime})\tau_x\Sigma_c(t^{\prime},t)
\right]
\\
Tr\left[\tau_x\left(G_{mf}(\Sigma_c \tau_x + \tau_x
\Sigma_c)G_{mf}\right)(t,t^{\prime})\Sigma_c(t^{\prime},t)
\right]&Tr\left[\tau_x\left(G_{mf}(\Sigma_c \tau_x + \tau_x
\Sigma_c)G_{mf}\right)(t,t^{\prime})\tau_x\Sigma_c(t^{\prime},t)
\right]
\end{pmatrix}
\end{eqnarray}
and $\delta\Pi^{\prime(1)}_q= {\cal O}\left(\frac{1}{N^2}\right)$
and therefore will not play a role in the subsequent discussion.

Integrating out the bosonic fields, and keeping terms upto ${\cal O}(1/N, \bar{b}_q)$ we get,
\begin{eqnarray}
S_K = i N b_{sp}\bar{b}_q Tr\left[ G_{mf}\{\tau_x \Sigma_c +
\Sigma_c \tau_x\}\right] -4 \lambda_{cl}b_{sp}\bar{b}_q -i b_{sp}
\bar{b}_qTr\left[D\delta\Pi^{\prime}_q \right]
\end{eqnarray}
Substituting for $\delta \Pi_q^{\prime}$, to ${\cal O}(1/N) $, the above becomes,
\begin{eqnarray}
&&S_K = i N b_{sp}\bar{b}_q Tr\left[ G_{mf}\{\tau_x \Sigma_c +
\Sigma_c \tau_x\}\right] -4 \lambda_{cl}b_{sp}\bar{b}_q 
\nonumber \\
&&- N b_{sp}\bar{b}_{q} \int \frac{d\epsilon}{2\pi} \int \frac{d\omega}{2\pi}
D^R(\epsilon) G^R_{mf}(\epsilon + \omega) \Sigma^K(\epsilon + \omega) G^R_{mf}(\epsilon + \omega)
\Sigma^K_c(\omega)
\end{eqnarray}

Thus the saddle point equation for $\epsilon_F$ (obtained from $\frac{\delta S_K}{\delta \bar{b}_q}=0$) 
reduces to
\begin{eqnarray}
&&\epsilon_F -E_0 + \frac{N\Gamma}{\pi} \sum_a \ln \frac{|\mu_a -\epsilon_F|}{D} = 
\label{NcorrR}\\
&& \frac{N\Gamma^2}{\pi^2 T_A}\left[ \int_{-D}^0d\omega 
\frac{\left[\sum_{a,b=L,R}
\frac{\Gamma_a \Gamma_b}{\Gamma^2}\left(\frac{1}{1-\frac{\omega +\mu_a}{\epsilon_F}}
-\frac{1}{1-\frac{\mu_b}{\epsilon_F}} \right)\right]}
{\omega -\frac{N\Gamma}{\pi}\sum_{a=L,R}\frac{\Gamma_a}{\Gamma} 
\ln{|1-\frac{\omega}{\epsilon_F -\mu_a}|}} 
+ 
\sum_{a\neq b}\int_0^{\mu_a -\mu_b} d\omega 
\frac{\left[
\frac{\Gamma_a \Gamma_b}{\Gamma^2}\left(\frac{1}{1-\frac{\omega +\mu_b}{\epsilon_F}}
-\frac{1}{1-\frac{\mu_a}{\epsilon_F}} \right)\right]}
{\omega -\frac{N\Gamma}{\pi}\sum_{a=L,R}\frac{\Gamma_a}{\Gamma} 
\ln{|1-\frac{\omega}{\epsilon_F -\mu_a}|}}
\right] \nonumber 
\end{eqnarray}

The above equation may be used to extract the ${\cal O}(1/N)$ correction to the saddle point
expression for the level energy $\epsilon_F$.
Writing
\begin{equation}
\epsilon_F = T_A + \frac{\beta}{N} \label{Ncorrepsilon}
\end{equation}
$T_A$ is given by Eq.~\ref{sp22a}, whereas from Eq.~\ref{NcorrR}, we get
\begin{eqnarray}
&&\beta = \frac{m^2 T_A}{1 + m_V}\sum_{a,b} \frac{\Gamma_a\Gamma_b}{\Gamma^2}
\left[ \int_{-(\mu_a-\mu_b)/T_A}^{D/T_A} dx 
\frac{\left(\frac{1}{1-\mu_a/T_A} -\frac{1}{1+x-\mu_b/T_A}\right)}
{1 + m \sum_i \frac{\Gamma_i}{\Gamma}\ln|1 + \frac{x}{1 -\mu_i/T_A}|}\right]
\label{betadef}
\end{eqnarray}
with $m_V$ defined in Eq.~\ref{muV}.

Using Eq.~\ref{Ldef} we may write
\begin{eqnarray}
\beta = \frac{m^2 T_A}{1 + m_V}\sum_{a,b} L_1^{a,b} =  
\frac{m^2 T_A}{1 + m_V}L_1
\label{betadef2}
\end{eqnarray}

Thus the two main results of this section is the ${\cal O}(1/N)$ corrections to the
level broadening ($b_{sp}^2$) and level position ($E_0 + \lambda$) which 
are given in Eq.~\ref{NcorrGam2} and Eq.~\ref{betadef2} respectively. These results will
be used in subsequent sections for the  
evaluation of various observables to ${\cal O}\left(\frac{1}{N}\right)$.

\section{Evaluation of $n_F$ to ${\cal O}\left(\frac{1}{N}\right)$} \label{nF}

\begin{figure}
\includegraphics[totalheight=3cm,width=6cm]{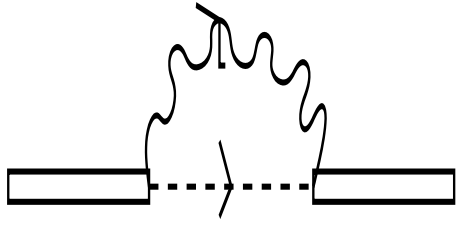}
\caption{Diagram contributing to the $1/N$ correction to $n_F$.
}
\label{fig3}
\end{figure}

In this section we will evaluate the local charge-density $n_F$ to ${\cal O}(1/N)$.
$n_F$ is given by 
\begin{eqnarray}
n_F= \sum_m\langle f_m^{\dagger} f_m\rangle = \frac{N}{2} \left[ 1 - i G^K_f\right]
\end{eqnarray} 
Thus we need to evaluate $G^K_f$ to
${\cal O}(1/N)$. For this we start by writing the Dyson equation for the fermionic 
Green's function correct to one loop, 
\begin{eqnarray}
G_f = G_{mf} + G_{mf} \Sigma_F G_{mf} \label{Gfpert}
\end{eqnarray}
where the second term in the above equation 
corresponds to the diagram in Fig~\ref{fig3}, and the  
$\Sigma_F$ are defined in Eq.~\ref{SigF1},~\ref{SigF2}.
The Keldysh component of Eq.~\ref{Gfpert} gives,
\begin{eqnarray}
G^K_f = G^K_{mf} + G^R_{mf} \Sigma^R_F G^K_{mf}  +G^R_{mf} \Sigma^K_F G^A_{mf} 
+G^K_{mf} \Sigma^A_F G^A_{mf}
\end{eqnarray}
We rewrite
\begin{equation}
n_F = n_F^0 + n_F^a + n_F^b  
\end{equation}
where
\begin{eqnarray}
n_F^0 &&= \frac{N}{2} \left[ 1 - i G^K_{mf}\right] \label{nF0}\\
&&= \frac{N \tilde{\Gamma}}{\pi\epsilon_F}\sum_{a=L,R}
\frac{\Gamma_a/\Gamma}{1 - \mu_a/\epsilon_F}  \\
n_F^a &&= \frac{-i N}{2} Tr\left[G^R_{mf} \Sigma^R_F G^K_{mf}+  G^K_{mf} \Sigma^A_F G^A_{mf}  \right]\\
n_F^b &&= \frac{-i N}{2} Tr\left[G^R_{mf} \Sigma^K_F G^A_{mf}  \right] 
\end{eqnarray}
We use Eqns~\ref{NcorrGam2},~\ref{Ncorrepsilon} 
and~\ref{betadef2} to correct ${\tilde \Gamma}/\Gamma, \epsilon_F$ to ${\cal O}(1/N)$ 
in Eq.~\ref{nF0} to obtain,
\begin{eqnarray}
n_F^0 = \frac{m_V}{1 + m_V}
\left[1 - \frac{1}{N} \frac{m^2}{1 + m_V} \sum_{a,b}
\left(L_2^{a,b} - \mu k^{a,b} + \frac{m/m_V}{1+ m_V} 
\left(\sum_i\frac{\Gamma_i/\Gamma}{(1 - \mu_i/T_A)^2}\right)L_1^{a,b} + (1 + m_V) I^{a,b}\right) 
\right]
\end{eqnarray}

Moreover using Eqn~\ref{SigF1},~\ref{SigF2}, one finds,
\begin{eqnarray}
n_F^a = \frac{1}{N}\frac{m^2}{1 + m_V} \sum_{a,b} L_2^{a,b} \label{nFa} \\
n_F^b = \frac{1}{N} \frac{m^2 m_V}{1 + m_V} \sum_{a,b} I^{a,b} - \frac{1}{N}
\frac{m^3}{1 + m_V} \sum_{a,b} k^{a,b} \label{nFb}
\end{eqnarray}

Adding Eqns~\ref{nF0},~\ref{nFa} and ~\ref{nFb} gives
\begin{eqnarray}
n_F = \frac{m_V}{1 + m_V} \left[1 + \frac{1}{N}\sum_{a,b}
\left(\frac{m^2/m_V}{1 + m_V}L_2^{a,b} - \frac{m^3/m_V}{1 + m_V}k^{a,b} 
 - \frac{m^3/m_V}{(1 + m_V)^2}L_1^{a,b} \sum_i \frac{\Gamma_i/\Gamma}{
\left(1 - \mu_i/T_A \right)^2}\right) \right] \label{nFtot}
\end{eqnarray}

Comparing Eq.~\ref{nFtot} with Eq.~\ref{NcorrGam1} we may write the expressions for $\frac{\tilde{\Gamma}}{\Gamma}$
in the following compact form,
\begin{eqnarray}
\frac{\tilde{\Gamma}}{\Gamma} = 1 - n_F - \frac{1}{N}\frac{m^2}{1 + m_V} \sum_{a,b} I^{a,b}
\end{eqnarray}
Upto ${\cal O}(1/N)$, above may be rewritten as
\begin{eqnarray}
\frac{\tilde{\Gamma}}{\Gamma} = 1 - n_F - \left(1-n_F\right)\frac{m^2}{N} \sum_{a,b} I^{a,b}
\label{gammfin}
\end{eqnarray}

The $I_{a,b}$ contain the divergent terms. If $\Lambda$ is an infrared cut-off,
and defining $V = |\mu_L -\mu_R|$, we find,
\begin{eqnarray}
&&\frac{\tilde{\Gamma}}{\Gamma} \simeq  
1 - n_F - \frac{1}{N}\frac{m^2}{1 + m_V}\frac{1}{\left(1+m_V\right)^2} 
\left(-\frac{\Gamma_L^2}{\Gamma^2}\frac{1}{\left(1-\mu_L/T_A\right)^2}
\ln\Lambda - \frac{\Gamma_R^2}{\Gamma^2}\frac{1}{\left(1-\mu_R/T_A\right)^2}
\ln\Lambda \label{gammfin1} \right.\\ 
&&\left.-\frac{2\Gamma_L \Gamma_R}{\Gamma^2}\frac{1}{\left(1-\mu_L/T_A \right)\left(1-\mu_R/T_A\right)}
\ln\frac{V}{T_A} + \frac{2\Gamma_L \Gamma_R}{\Gamma^2}
\frac{1}{\left(1-\mu_L/T_A \right)\left(1-\mu_R/T_A\right)}
\frac{V}{T_A} \int_0^{V/T_A} \frac{dx}{x^2}
\right)
\nonumber 
\end{eqnarray}
The above expression will be useful in the next section when we study the bosonic correlation
function.

\section{Bosonic correlation function: Decay due to current induced decoherence} \label{Xray}

The full bosonic correlation function (combining both saddle point and fluctuation corrections) is
\begin{eqnarray}
&&\bar{D}^K(t,t^{\prime}) = -i\langle{\{b_{sp}(t) + \delta b(t), 
b_{sp} + \delta b^{\dagger}(t^{\prime}) \}}\rangle = -2i\frac{\tilde{\Gamma}}{\Gamma} +
D^K_{b,b^*}(t,t^{\prime})  
\end{eqnarray}
where $D^K_{b,b^*} = -i\langle\{\delta b(t), \delta b^{\dagger}(t^{\prime})\}\rangle$, and its expression in
frequency space is given in Eq.~\ref{DKdef1}. 
Using Eq.~\ref{DKdef1}
\begin{eqnarray}
D^K_{b,b^*}(t) = \int_{-\infty}^{\infty} \frac{d\Omega}{2\pi} e^{-i \Omega t} D^K_{b,b^*}(\Omega) = \frac{-i}{N} 
\frac{m^2}{1 + m_V} \sum_{a,b=L,R} \frac{\Gamma_a \Gamma_b}{\Gamma^2}\gamma_{ab}(t)
\end{eqnarray}
where
\begin{eqnarray}
\gamma_{ab}(t) = \int_{-\infty}^{\infty} d\Omega e^{-i\Omega t} sgn\left(\Omega + \mu_a -\mu_b \right)
\frac{\frac{\Omega + \mu_a - \mu_b}{(1- \frac{\Omega + \mu_a}{T_A})(1-\frac{\mu_b}{T_A})}}
{\left(\Omega -\frac{N \Gamma}{\pi} \sum_i \frac{\Gamma_i}{\Gamma}\ln|1 - \frac{\Omega}{T_A-\mu_i}|\right)^2}
\end{eqnarray}
where in the long time limit,
\begin{eqnarray}
&&\gamma_{aa}(t) = \frac{2}{\left(1-\frac{\mu_a}{T_A} \right)^2 \left(1 + m_V \right)^2}
\int_0^{\infty} d\Omega \frac{1}{\Omega}\cos{\Omega t}  \nonumber \\
&&\simeq \frac{-2}{\left(1-\frac{\mu_a}{T_A} \right)^2 \left(1 + m_V \right)^2}
\ln{\left(\Lambda t T_A\right)} 
\end{eqnarray}
$\Lambda$ is a cutoff introduced to take care of the infra-red divergences. This term, as we shall show
will be canceled by the corresponding infrared divergence from Eq.~\ref{NcorrGam2}.

Similarly one finds (for $V = |\mu_L-\mu_R|$),
\begin{eqnarray}
\gamma_{LR} + \gamma_{RL} = \frac{2}{\left(1-\frac{\mu_L}{T_A}\right) 
\left(1 - \frac{\mu_R}{T_A} \right) \left(1+m_V\right)^2} \left[2\int_{V}^{\infty} 
\frac{d\Omega}{\Omega} \cos{\Omega t}  + 2 V \int_0^{V}\frac{d\Omega}{\Omega^2}\cos{\Omega t}\right]
\end{eqnarray}

Therefore the full bosonic correlation function is 
\begin{eqnarray}
&&\bar{D}^K(t) =\nonumber\\
&& \simeq -2i\left[\frac{\tilde{\Gamma}}{\Gamma}  + \frac{1}{N} \frac{m^2}{1+m_V}\frac{1}
{\left(1 + m_V\right)^2}
\left( 
-\frac{\Gamma^2_L}{\Gamma^2}\frac{1}{\left(1-\mu_L/T_A\right)^2}\ln{\left(\Lambda t T_A\right)} 
 - \frac{\Gamma^2_R}{\Gamma^2}\frac{1}{\left(1-\mu_R/T_A\right)^2}\ln{\left(\Lambda t T_A\right)}  
\right. \right. \label{DKfull}\\
&&\left. \left. 
+ \frac{2 \Gamma_L \Gamma_R}{\Gamma^2} \frac{1}{\left(1-\mu_L/T_A\right)
\left(1 - \mu_R/T_A\right)} \int_{V}^{\infty} 
\frac{d\Omega}{\Omega} \cos{\Omega t}  + 
\frac{2 \Gamma_L \Gamma_R}{\Gamma^2} \frac{1}{\left(1-\mu_L/T_A\right) 
\left(1 - \mu_R/T_A\right)} V \int_0^{V}\frac{d\Omega}{\Omega^2}\cos{\Omega t}
 \right)\right]
\nonumber
\end{eqnarray}
Combining the above with expression for $\tilde{\Gamma}/\Gamma$ in Eq.~\ref{gammfin1}, one
finds that the infrared divergences cancel to give,
\begin{eqnarray}
&&\bar{D}^K(t) = -2i\left(1-n_F\right)\left[1 - \frac{1}{N}\frac{m^2}{\left(1 + m_V\right)^2} 
\left(\frac{\Gamma^2_L}{\Gamma^2}\frac{1}{\left(1-\mu_L/T_A\right)^2}\ln{\left(t T_A\right)}
+\frac{\Gamma^2_R}{\Gamma^2}\frac{1}{\left(1-\mu_R/T_A\right)^2}\ln{\left(t T_A\right)} \right.\right. \nonumber \\
&&\left. \left. - \frac{2 \Gamma_L \Gamma_R}{\Gamma^2}\frac{1}{\left(1-\mu_L/T_A\right)
\left(1 - \mu_R/T_A\right)} \left(\ln\frac{V}{T_A}+\int_V^{\infty}\frac{d\Omega}{\Omega}\cos\Omega t\right)
\right.\right.\nonumber \\
&&\left.\left.-\frac{2 \Gamma_L \Gamma_R}{\Gamma^2}\frac{1}{\left(1-\mu_L/T_A\right)
\left(1 - \mu_R/T_A\right)} \frac{V}{T_A}\int_0^{V/T_A} \frac{dx}{x^2} \left(\cos{xt T_A}-1\right)
\right)
\right] \label{DKtot}
\end{eqnarray}
Note that the above expression is correct to ${\cal O}\left(1/N\right)$. Therefore $1-n_F$ needs to be computed
only to the saddle point level for all terms except the first term in the square brackets. 

For long times $Vt\gg 1$, Eq.~\ref{DKtot} reduces to
\begin{eqnarray}
\bar{D}^K(t) \simeq -2i\left(1-n_F\right)\left[1 - \frac{c_L}{N}\ln{tT_A} -\frac{c_R}{N}\ln{tT_A}
-\frac{c_{dec}}{N} V t\right]
\label{DKlongtime}
\end{eqnarray}
where
\begin{eqnarray}
c_{L,R} = \frac{m^2}{\left(1 + m_V\right)^2} 
\frac{\Gamma^2_{L,R}}{\Gamma^2}\frac{1}{\left(1-\mu_{L,R}/T_A\right)^2}  \label{cL}\\
c_{dec} =  \frac{m^2}{\left(1 + m_V\right)^2}
\frac{2 \Gamma_L \Gamma_R}{\Gamma^2}\frac{1}{\left(1-\mu_L/T_A\right)
\left(1 - \mu_R/T_A\right)}  \label{cdec}
\end{eqnarray}

If one were to compute the correlation function to higher orders in $\frac{1}{N}$, Eq.~\ref{DKlongtime}
signals the following behavior 
\begin{eqnarray}
\bar{D}_K(t) \sim -2i \left(1-n_F\right) \exp \left[-\frac{1}{N}\sum_{i=L,R}c_{i}
\ln t T_A - \frac{c_{dec}}{N} V t\right]
\end{eqnarray}
Thus the slow power-law decay in time in equilibrium of the bosonic correlator is replaced by a 
rapid exponential decay at non-zero voltages whose origin is current induced decoherence. Each
of the exponents $c_{i,dec}$ is consistent with what one might expect from nonequilibrium
X-ray edge physics~\cite{Ng}.

The above decoherence rate appearing in the bosonic correlation function has consequences for
physical observables such as the susceptibility and the conductance
which we evaluate in subsequent sections. 

\section{Evaluation of susceptibility to ${\cal O}\left(\frac{1}{N}\right)$} \label{susc}

\begin{figure}
\includegraphics[totalheight=4cm,width=4cm]{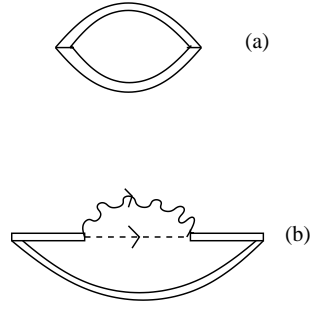}
\caption{Diagrams contributing to the susceptibility to ${\cal O}(1/N)$.
}
\label{fig4}
\end{figure}

The spin response function is given by
\begin{equation}
\chi^R(\Omega) = \frac{i}{2} \sum_{m}\left(g\mu_B m \right)^2 \int \frac{d\omega}{2\pi}
\left[G^R_f(\omega + \Omega)G^K_f(\omega) + G^K_f(\omega + \Omega) G^A_f(\omega)\right] \label{chiR1}
\end{equation}

To ${\cal O }(1/N)$, from Eq.~\ref{chiR1}
\begin{eqnarray}
\chi^R(\Omega=0) = \chi_0^R + \chi_1^R + \chi_2^R
\end{eqnarray}
where $\chi_0^R$ is the saddle point expression for the susceptibility (diagram $(a)$
in Fig~\ref{fig4}) with the level energy and
broadening corrected to ${\cal O}(1/N)$, while $\chi_{1,2}^R$ arise due to corrections to the electronic
Green's functions to one-loop (diagram $(b)$ in Fig~\ref{fig4}). In particular,
\begin{eqnarray}
\chi^R_1 = i\sum_{m}\left(g \mu_B m\right)^2
Tr\left[G^R_{mf}\Sigma^R_fG^R_{mf} G^K_{mf}\right] \\
\chi^R_2 = i\sum_{m}\left(g \mu_B m\right)^2
Tr\left[\{G^R_{mf} (\Sigma^R_f G^K_{mf} + \Sigma^K_f G^A_{mf}) + G^K_{mf} \Sigma^A_f G^A_{mf} \}  G^R_{mf}\right]
\end{eqnarray}

In order to compute $\chi^R_0$, we use Eq.~\ref{chisp} and correct for $\tilde{\Gamma}/\Gamma$ using 
Eq.~\ref{NcorrGam2}, and correct for $T_A$ using Eq.~\ref{betadef2}. We find,
\begin{eqnarray}
\chi_0^R = \chi_{sp}\left[1 + \frac{1}{N}\left(-\frac{m^2}{1 + m_V}L_2 
+ \frac{m^3}{1 + m_V}k + \left[-2\frac{m^2}{1 + m_V} \frac{S_3}{S^2_2}+
\frac{m^3}{(1 + m_V)^2}\right]L_1 S_2 - m^2 I \right) \right]\label{chi0}
\end{eqnarray}

Using Eq.~\ref{SigF1} it is straightforward to show
\begin{eqnarray}
\chi^R_1 = \left(\frac{g^2 \mu_B^2}{3} J(J+1)\right)\frac{1}{T_A}\frac{4}{3} \frac{1}{N} 
\left(\frac{m^2}{1 + m_V}\right)  L_3 = \chi^R_{sp} \frac{2}{3}\frac{1}{N}\frac{m L_3}{S_2}
\label{chi1}
\end{eqnarray}
whereas
\begin{eqnarray}
&&\chi^R_2 = 2 \chi^R_1 + \\
&&\frac{g^2 \mu_B^2}{T_A} \frac{J(J+1)}{3}\frac{1}{N} 
\frac{m^3}{1 + m_V} \sum_{ab}\frac{\Gamma_a \Gamma_b}{\Gamma^2}
\int_{-(\mu_a-\mu_b)/T_A}^{D/T_A} dx 
\frac{\left(\frac{1}{1-\mu_a/T_A} - 
\frac{1}{1 + x - \mu_b/T_A}\right)}{\left(x + m \sum_i
\frac{\Gamma_i}{\Gamma} \ln{|1 + \frac{x}{1-\mu_i/T_A}|} \right)^2} \sum_i \frac{\Gamma_i/\Gamma}
{\left(1 + x - \mu_i/T_A \right)^2} \nonumber 
\end{eqnarray}
Collecting all the terms together 
\begin{eqnarray}
\chi^R = \chi_{sp} \left[1 + \frac{1}{N}\left(-\frac{m^2}{1 + m_V}L_2 + 
\frac{m^3}{1+m_V}k + \frac{m^3}{(1 + m_V)^2}L_1 S_2 - 2 \frac{S_3}{S_2}\frac{m^2}{1+m_V}L_1
+ 2 m L_3 S_2 - m^2 I + F \right) \right]
\end{eqnarray}
where
\begin{eqnarray}
F = \frac{m^2}{S}\sum_{a,b}  \frac{\Gamma_a \Gamma_b}{\Gamma^2}
\int_{-(\mu_a-\mu_b)/T_A}^{D/T_A} dx 
\frac{\left(\frac{1}{1-\mu_a/T_A} - 
\frac{1}{1 + x - \mu_b/T_A}\right)}{\left(x + m \sum_i
\frac{\Gamma_i}{\Gamma} \ln{|1 + \frac{x}{1-\mu_i/T_A}|} \right)^2} \sum_i \frac{\Gamma_i/\Gamma}
{\left(1 + x - \mu_i/T_A \right)^2}
\end{eqnarray}

Now it can be shown that
\begin{eqnarray}
\frac{m^3}{1+m_V}k -m^2 I + F = -\frac{m^2}{1 + m_V}M
\end{eqnarray}
where
\begin{eqnarray}
&&M = \sum_{a,b} M^{a,b}\\
&&M^{a,b} = \frac{\Gamma_a \Gamma_b}{\Gamma^2} \int_{-(\mu_a -\mu_b)/T_A}^{D/T_A} dx
\frac{\left(\frac{x +  (\mu_a -\mu_b)/T_A}{(1-\mu_a/T_A)(1+x -\mu_b/T_A)}\right) 
}{(x + m \sum_{i}\frac{\Gamma_i}{\Gamma} 
\ln{|1+ \frac{x}{1-\mu_i/T_A}|})^2}  \nonumber \\
&&\sum_i \frac{\Gamma_i}{\Gamma}\left[ \left(\frac{1 + m_V}{S} \right)\frac{x^2 + 2 x \left(1-\mu_i/T_A \right)}
{(1+x-\mu_i/T_A)^2 (1-\mu_i/T_A)^2}
-\frac{m x}{(1+x-\mu_i/T_A) (1-\mu_i/T_A)}\right]\label{Mdef}
\end{eqnarray}

Therefore the static spin-susceptibility becomes,
\begin{eqnarray}
\chi^R = \chi_{sp} \left[1 + \frac{1}{N}\left(-\frac{m^2}{1 + m_V}L_2 
+ \frac{m^3}{(1 + m_V)^2}L_1 S_2 - 2 \frac{S_3}{S_2}\frac{m^2}{1+m_V}L_1
+ 2 m \frac{L_3}{S_2} - \frac{m^2}{1 + m_V} M\right) \right] \label{stsusc}
\end{eqnarray}
with $\chi_{sp}$ given in Eq.~\ref{chisp1}. 
Eq.~\ref{stsusc} may be expanded in powers of $\mu_{L,R}/T_A$. 
In particular in the Kondo limit ($n_F \rightarrow 1$ or $m_V \gg 1$), Eq.~\ref{stsusc} is found to have the
form, 
\begin{eqnarray}
&&\chi_S^{n_F=1} = \frac{g^2\mu_B^2 J(J+1)}{3 T_A^0}
\left[1 + 1.5 \frac{\Gamma_L \Gamma_R}{\Gamma^2}\left(\frac{\mu_L -\mu_R}{T_A^0}\right)^2\right]
\left[1 + \frac{m}{N}\left(-L_2^{eq} - L_1^{eq} + 2 L_3^{eq} - m J_0^{eq}\right)
+\frac{1}{N}\left(\sum_{i=L,R}\frac{\Gamma_i \mu_i}{\Gamma T^0_A}\right) C_{S1} +
\right.\nonumber \\ 
&&\left. \frac{1}{N}\left(\sum_{i=L,R}\frac{\Gamma_i \mu_i^2}{\Gamma (T^0_A)^2}\right) C_{S2} + 
\frac{1}{N}\left(\sum_{i=L,R}\frac{\Gamma_i \mu_i}{\Gamma T^0_A}\right)^2 \left(C_{S3} -C_{S1}\right)
-\frac{4.5}{N}\frac{\Gamma_L\Gamma_R}{\Gamma^2}\left(\frac{\mu_L -\mu_R}{T_A^0}\right)^2 
\right] \label{neqsusc}
\end{eqnarray}
The expressions for the $C_{Si}$ have been given in Appendix~\ref{Csi}, and may be evaluated numerically. 
The $L_p^{eq}$ are defined in Eq.~\ref{Lpeq}, and
\begin{eqnarray}
J_0^{eq} = \int_{0}^{D/T_A} \frac{dx}{\left(x+m \ln\left(1+x\right)\right)^2}
\frac{x^2}{\left(1+x\right)^3}
\end{eqnarray}
Let us assume that the chemical potential of the left lead $\mu_L = V/2$, and that for the right lead is
$\mu_R = -V/2$. Let us define 
\begin{eqnarray}
C_{S0} = m\left(-L_2^{eq} - L_1^{eq} + 2 L_3^{eq} - m J_0^{eq}\right)
\end{eqnarray}
We now rewrite Eq.~\ref{neqsusc} as follows,
\begin{eqnarray}
&&\chi_S^{n_F=1} = \frac{g^2\mu_B^2 J(J+1)}{3 T_A^0}\left(1 + \frac{C_{S0}}{N}\right)
\left[1 + 1.5 \frac{\Gamma_L \Gamma_R}{\Gamma^2}\left(\frac{V}{T_A^0}\right)^2\left(1 + \frac{2 C_{S0}}{N}
-\frac{2 C_{S0}}{N}\right)
+\frac{1}{N}\left(\frac{\Gamma_L -\Gamma_R}{\Gamma}\right)\left(\frac{V}{2T^0_A}\right) C_{S1} +
\right.\nonumber \\ 
&&\left. \frac{1}{N}\left(\frac{V}{2T^0_A}\right)^2 C_{S2} + 
\frac{1}{N}\left(\frac{\Gamma_L -\Gamma_R}{\Gamma}\right)^2 
\left(\frac{V}{2 T^0_A}\right)^2 \left(C_{S3} -C_{S1}\right)
-\frac{4.5}{N}\frac{\Gamma_L\Gamma_R}{\Gamma^2}\left(\frac{V}{T_A^0}\right)^2 
\right] \label{neqsusc2}
\end{eqnarray}
where terms higher order than $\frac{1}{N}\frac{V^2}{(T_A^0)^2}$ have been dropped. 
Defining the Kondo temperature to ${\cal O}(1/N)$ as~\cite{Read87}
\begin{eqnarray}
T_K = T_A^0\left(1-\frac{C_{S0}}{N}\right) \label{TKNdef}
\end{eqnarray}
Eq.~\ref{neqsusc} can be recast in the following universal form
\begin{eqnarray}
&&\chi_S^{n_F=1} = \frac{g^2\mu_B^2 J(J+1)}{3 T_K}
\left[1 + 1.5 \frac{\Gamma_L \Gamma_R}{\Gamma^2}\left(\frac{V}{T_K}\right)^2
+\frac{1}{N}\left(\frac{\Gamma_L -\Gamma_R}{\Gamma}\right)\left(\frac{V}{2T_K}\right) C_{S1} +
\right.\nonumber \\ 
&&\left. \frac{1}{N}\left(\frac{V}{2T_K}\right)^2 \left(C_{S2} + C_{S3} -C_{S1}\right)  
-\frac{1}{N}\left(4.5+3C_{S0} + C_{S3}-C_{S1}\right)\frac{\Gamma_L\Gamma_R}{\Gamma^2}\left(\frac{V}{T_K}\right)^2 
\right] \label{neqsusc3}
\end{eqnarray}
In the 
Kondo limit $m\gg 1$, the coefficients in the above
equation take the following universal values $C_{S1} \rightarrow 0.01$,
$ \left(C_{S2} + C_{S3} -C_{S1}\right) \rightarrow -0.005$,
$\left(3C_{S0} + C_{S3}-C_{S1}\right)\rightarrow -4.94$. 

\section{Evaluation of the spectral density and conductance to ${\cal O}\left(\frac{1}{N^2}\right)$} \label{spect}

\begin{figure}
\includegraphics[totalheight=5cm,width=8cm]{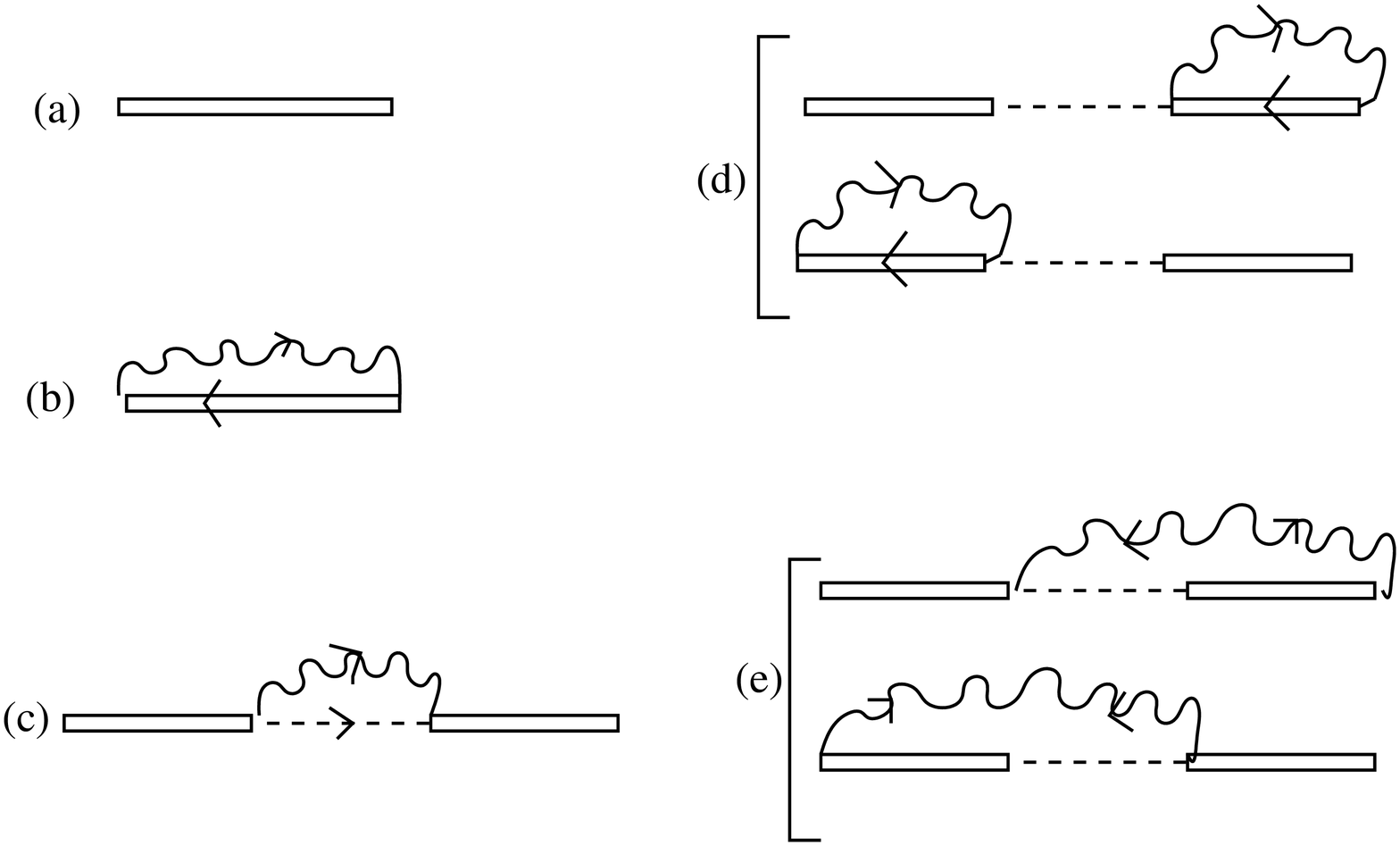}
\caption{Diagrams needed for the computation of the impurity spectral density to
${\cal O}(1/N^2)$ 
and hence the conductance to ${\cal O}\left(1/N^2\right)$
}
\label{fig6}
\end{figure}

The retarded Green's function, whose imaginary part gives the impurity spectral density is
\begin{eqnarray}
&&G^R_{f,b}\left(t,t^{\prime}\right) = -i 
T\langle b_-^{\dagger}(t) f_-(t) f_-^{\dagger}(t^{\prime}) b_-({t^{\prime}})\rangle 
- i \langle f_+^{\dagger}(t^{\prime})b_+(t^{\prime}) b_-^{\dagger}(t) f_-(t)\rangle\\
&&= i\left[G^f_{--}(t,t^{\prime}) D^{b,b^*}_{--}(t^{\prime},t) - G^f_{-+}(t,t^{\prime}) D^{b,b^*}_{+-}(t^{\prime},t) 
\right]\label{speca}
\end{eqnarray}
We are only interested in evaluating the imaginary part of Eq.~\ref{speca} which at leading order
(saddle point level) is ${\cal O}\left(\frac{1}{N}\right)$. 
There are five diagrams that contribute to the above expression to ${\cal O}\left(\frac{1}{N^2}\right)$ which are
shown in Fig~\ref{fig6}. 
For convenience we write
\begin{eqnarray}
Im{\left[G^R_{f_m,b}\right]} = T_a + T_b + T_c + T_d + T_e   
\end{eqnarray}
where $T_i$ is the contribution from the $i$-th diagram and corresponds to 
\begin{eqnarray}
&&T_a(\Omega) =\frac{-i\Gamma}{\left(T_A-\Omega\right)^2}\frac{1}{(1 + m_V)^2} 
\left[1 + \frac{2}{N} \left(-\frac{m^2}{1 + m_V}L_2 + \frac{m^3}{1+m_V}k 
+ \frac{m^3}{(1 + m_V)^2}L_1 S \right. \right. \nonumber \\
&&\left. \left. 
- \frac{m^2}{1 + m_V}\frac{L_1}{\left(1 - \frac{\Omega}{T_A}\right)}
- \mu^2 I \right)\right]\label{T1} \\
&&T_b(\Omega) = Im{\left[\frac{i}{2}\int \frac{d\omega}{2\pi} \{
G^K_{mf}(\omega + \Omega) D^A_{b,b^*}(\omega) + G^R_{mf}(\omega + \Omega)D^K_{b,b^*}(\omega)
\}\right]} \label{T2} \\
&&T_c(\Omega) = \frac{\tilde{\Gamma}}{\Gamma}Im{\left[\left(G^R_{mf}(\Omega)\right)^2 
\frac{i}{2}\int \frac{d\omega}{2\pi}\{D^R_{b,b^*}(\omega + \Omega)\Sigma^K_c(-\omega) + 
D^K_{b,b^*}(\omega + \Omega)
\Sigma^R_c(-\omega)\}\right]} \label{T3} \\
&&T_d(\Omega) = 2\frac{\tilde{\Gamma}}{\Gamma}Im{\left[G^R_{mf}(\Omega) \Sigma^R_c(\Omega)
\frac{i}{2}\int \frac{d\omega}{2\pi}\{G^K_{mf}(\omega + \Omega) D^A_{b,b^*}(\omega)
+G^R_{mf}(\omega + \Omega) D^K_{b,b^*}(\omega)\}\right]} \label{T4} \\
&&T_e(\Omega) = \frac{\tilde{\Gamma}}{\Gamma}Im\left[G^R_{mf}(\Omega)\frac{i}{2} 
\int \frac{d\omega}{2\pi}\left[D^R_{b,b}(\omega + \Omega)\{\left(\Sigma^R_c(-\omega) + \Sigma^A_c(-\omega) \right)
G^K_{mf}(-\omega) + \Sigma^K_c(-\omega)\left(G^R_{mf}(-\omega) 
+ G^A_{mf}(-\omega)\right)\} \nonumber \right. \right. \\
&& \left. \left. + D^K_{b,b}(\omega + \Omega)\Sigma^R_c(-\omega)G^R_{mf}(-\omega) \right]\right] \label{T5}
\end{eqnarray}
The above terms have been evaluated in Appendix~\ref{speccal}. 

We now present results for the conductance for the case of symmetric couplings to the 
leads ($\Gamma_L = \Gamma_R$)
and $\mu_L = V/2, \mu_R = -V/2$. Defining 
\begin{eqnarray}
G_0 = \frac{N e^2}{h}\frac{4 \Gamma_L \Gamma_R}{\Gamma^2}
\left(\frac{\pi}{N}\right)^2\left(\frac{m_0}{1 +m_0}\right)^2
\end{eqnarray}
and the functions
\begin{eqnarray}
&&k^0 = \int_{0}^{D/T_A} d x 
\frac{\left(\frac{x}{1+x}\right)^2}{(x + m  
\ln{(1+ x)})^2} \\
&&p^0 = \int_{0}^{D/T_A} d x 
\frac{\left(\frac{x}{1+x}\right)^2 \frac{(x^2 + 3x + 3)}{(1+x)^2}}{(x + m  
\ln{(1+ {x})})^2}\\
&&p^1 = \int_{0}^{D/T_A} d x 
\frac{\left(\frac{x}{1+x}\right)^2 \frac{(3x^2 + 8x + 6)}{(1+x)^2}}{(x + m  
\ln{(1+{x})})^2}\\
&&t_1= \frac{1}{2} \int_0^{D/T_A} \frac{dx}{\left((x + m  
\ln{(1+ x)}\right)^2} \frac{x(x+2)}{(1+x)^2} \left(1-\frac{1}{(1+x)^2}  \right) \\
&&t_2=  \int_0^{D/T_A} \frac{dx}{\left((x + m  
\ln{(1+ x)}\right)^3} \frac{x(x+2)}{(1+x)^2}  \frac{x^2}{(1+x)^2}   \\
&&t_3= \frac{1}{2} \int_0^{D/T_A} \frac{dx}{\left((x + m  
\ln{(1+ x)}\right)^2} \frac{x(x+2)}{(1+x)^2} \left(1-\frac{1}{1+x}  \right) 
\end{eqnarray}
we obtain the following expression for the conductance,
\begin{eqnarray}
\frac{G(V)}{G_0}
&&=1+\frac{2}{N}\frac{m_0}{1+m_0}\left[L^{eq}_2 -m_0 k^{0}- \frac{m_0}{1+m_0} L^{eq}_1 
\right]  -\frac{3V^2}{4(T^0_A)^2} \frac{m_0}{1+m_0} +
\label{G1Nneq}\\
&& +\frac{1}{N} \frac{V^2}{(T^0_A)^2}  \left[-\frac{m_0}{1+m_0}  + \frac{17}{2} 
\left(\frac{m_0}{1+m_0} \right)^2 
-6  \left(\frac{m_0}{1+m_0} \right)^3 +   
\frac{5}{4} \left(\frac{m_0}{1+m_0} \right)^4    \right] +             
\\
&& +\frac{1}{N} \frac{3V^2}{4(T^0_A)^2} 
\left[-m_0 t_1+m_0^2 t_2  +\frac{m_0^2}{1+m_0} t_3  \right]   \\
&&+ \frac{1}{N} \frac{V^2}{2(T^0_A)^2} \left[ \frac{m_0(m_0^2-7m_0-1/2)}{(1+m_0)^2}L^{eq}_2 
+\frac{3m_0^2(7m_0+2)}{2(1+m_0)^2}k^{0} + 
\frac{3m_0^2(m_0^2+3m_0-1)}{(1+m_0)^3} L^{eq}_1+ \right. \nonumber \\
&&\left. + \frac{m_0(2m_0^2+m_0+2)}{(1+m_0)^2}L^{eq}_3 -\frac{3m_0(2m_0-1)}{1+m_0}  
L^{eq}_4  
+\frac{3m_0^2(m_0-1)}{1+m_0} p^{0}-m_0^2p^{1}   \right]
\end{eqnarray}
Note that in equilibrium, the above equation reduces to~\cite{Read87} 
$G = \frac{N e^2}{h}\frac{4 \Gamma_L \Gamma_R}{\Gamma^2}
\left(\frac{\pi n_F}{N}\right)^2$ as expected from the Friedel sum rule.  

The expression for the conductance in the Kondo regime may be obtained by setting $m_0 \gg 1$ in the above equation.
We find
\begin{eqnarray}
G^{n_F=1}(V;\Gamma_L = \Gamma_R) =\frac{N e^2}{h}\frac{4 \Gamma_L \Gamma_R}{\Gamma^2}
\left(\frac{\pi}{N}\right)^2 \left[1 - \frac{3}{4}\left(\frac{V}{T_K}\right)^2
+ \frac{1}{2N}\left(\frac{V}{T_K}\right)^2\left(5.5 + C_{G1}\right)\right] \label{GneqNK}
\end{eqnarray}
with $T_K$ defined in Eq.~\ref{TKNdef}, and the coefficient $C_{G1}$ given in Eq.~\ref{CG1}. 
In the Kondo limit, $C_{G1} \rightarrow -2.77$.

\section{Conclusions} \label{conc}

In summary, we have presented results for the nonequilibrium infinite-U Anderson model
using Keldysh functional integral methods. The approach has been to use   
$1/N$ as the small parameter in the theory which allows us to develop
a systematic perturbation theory for the nonequilibrium problem. 
The results derived are
valid for an applied voltage small as compared to the Kondo temperature when the
effect of fluctuations are small. 
Physical quantities such as the impurity spin susceptibility and the conductance are
calculated to ${\cal O}\left(\frac{1}{N}\left(\frac{V}{T_K}\right)^2 \right)$.
The voltage expansions are found to 
show rich behavior by depending on 
different combinations of the couplings to the left and 
right leads such as: $\left(\frac{\Gamma_L -\Gamma_R}{\Gamma}\right)\frac{V}{T_K},
\frac{\Gamma_L^2 + \Gamma_R^2}{\Gamma^2}\left(\frac{V}{T_K}\right)^2,\frac{\Gamma_L \Gamma_R}{\Gamma^2}
\left(\frac{V}{T_K}\right)^2$.
While terms of the first kind give rise to rectification type behavior,
{\sl i.e.} $\chi_S(V) \neq \chi_S(-V)$ and $G(V) \neq G(-V)$, 
the last term is associated with current induced decoherence as it arises due to inelastic processes that can
occur in an energy window V. This term is also 
found to cause the bosonic correlation function to decay rapidly in time.  
The approach developed in this paper is rather general, and therefore may be easily adaptable to a variety of
out-of-equilibrium systems.

An interesting question that arises is to what extent the results obtained in this paper are also
valid for $N=2$. It is known for equilibrium systems that a naive extrapolation 
of the results of large-N to $N=2$ when compared with exact Bethe-Ansatz
results not only give incorrect  numerical values of various quantities (such as the Wilson
ratio and the zero-bias conductance), but also make qualitatively incorrect predictions
for the temperature dependence of observables. Precisely how the extrapolation goes
wrong has been discussed in Appendix~\ref{extrap}. On the
other hand, comparison with exact results~\cite{Bickers87}
reveal that large-N works very well for $N \geq 4$. However, one of the results of this
paper has been the observation that $G(-V) \neq G(V), \chi_S(V) \neq \chi_S(-V)$ for
unequal coupling to the two leads. This asymmetry is rather generic and
will exist whenever the system is away from particle-hole symmetry and therefore should be
observed for the nonequilibrium $N=2$ Anderson model away
from the particle-hole symmetry point $E_0 = -\frac{U}{2}$. However,
for a small voltage expansion of the conductance for $N=2$ we do not expect the appearance
of a linear in voltage term as found in Eq.~\ref{Gspsumm}. This is because the $N=2$ case 
has a maximal conductance per channel of $e^2/h$, and such a linear term would imply that the conductance
can become larger than this value, which is unphysical. 
An asymmetry can
very well appear at cubic order ($~\frac{\Gamma_L-\Gamma_R}{\Gamma}
\left(\frac{V}{T_K}\right)^3$) in the small voltage expansion of $G$. Note that for $N \gg 1$,
the conductance per channel is a small number of ${\cal O}\left(1/N^2\right)$. 
Therefore for this case a linear 
in voltage term in the conductance does not violate unitarity.

{\it Acknowledgments}: AM gratefully acknowledges helpful
discussions with Natan Andrei, Piers Coleman and Andrew Millis. 
This work was supported by NSF-DMR 0705584.

\appendix

\section{Evaluation of bosonic self-energies and propagators} \label{Pieval}

In this section we evaluate explicit expressions for $\Pi^{R,A,K}$ defined in Eq.~\ref{Pidef}.
In particular
\begin{eqnarray}
\Pi^R(1,2)  = \frac{-i N}{2} \left[G^R_{mf}(1,2) \Sigma^K_c(2,1) + G^K_{mf}(1,2)\Sigma^A_c(2,1) \right]\\
\Pi^A(1,2) = \frac{-i N}{2} \left[G^A_{mf}(1,2) \Sigma^K_c(2,1) + G^K_{mf}(1,2)\Sigma^R_c(2,1) \right] \\
\Pi^K(1,2) = \frac{-i N}{2} \left[G^R_{mf}(1,2) \Sigma^A_c(2,1) + G^A_{mf}(1,2)\Sigma^R_c(2,1)
+ G^K_{mf}(1,2)\Sigma^K_c(2,1)\right]
\end{eqnarray}
The above may be easily evaluated.
We obtain,
\begin{eqnarray}
\Pi_R(\Omega) = \Pi^{\prime}_R(\Omega) + i \Pi^{\prime \prime}_R(\Omega)
\end{eqnarray}
where
\begin{eqnarray}
\Pi^{\prime}_R(\Omega) &&= \frac{N\Gamma}{\pi} \left[ \frac{\Gamma_L}{\Gamma}
\ln\frac{\sqrt{(\Omega + \mu_L - E_0 -\lambda_{cl})^2 + \tilde{\Gamma}^2}}{D}
+ \frac{\Gamma_R}{\Gamma}\ln\frac{\sqrt{(\Omega + \mu_R - E_0 -\lambda_{cl})^2 + \tilde{\Gamma}^2}}{D}   \right]\\
\Pi^{\prime \prime}_R(\Omega) &&= \frac{-N\Gamma}{\pi}  
\left[\frac{\Gamma_L}{\Gamma}\left(
\arctan\frac{\tilde{\Gamma}}{\mu_L - E_0 -\lambda_{cl}} -
\arctan\frac{\tilde{\Gamma}}{\Omega + \mu_L - E_0 -\lambda_{cl}}
\right) + \right. \\
&&\left. \frac{\Gamma_R}{\Gamma}\left(
\arctan\frac{\tilde{\Gamma}}{\mu_R - E_0 -\lambda_{cl}} -
\arctan\frac{\tilde{\Gamma}}{\Omega + \mu_R - E_0 -\lambda_{cl}}
\right)   \right]\nonumber
\end{eqnarray}
Defining $\lambda_{cl} + E_0 = \epsilon_F$, and to ${\cal O}(1/N)$, the above expressions simplify to
\begin{eqnarray}
&&\Pi^{\prime}_R(\Omega) = \frac{N\Gamma}{\pi} \left[ \frac{\Gamma_L}{\Gamma}
\ln\frac{|\Omega + \mu_L - \epsilon_F|}{D}
+ \frac{\Gamma_R}{\Gamma}\ln\frac{|\Omega + \mu_R - \epsilon_F|}{D}   \right] \label{PiRReN}\\
&&\Pi^{\prime \prime}_R(\Omega) = \frac{-N\Gamma\tilde{\Gamma}}{\pi \epsilon_F}
\left[\frac{\Gamma_L}{\Gamma}\left(\frac{1}{1 - \frac{\Omega + \mu_L}{\epsilon_F}} -
\frac{1}{1 - \frac{\mu_L}{\epsilon_F}}
\right) + \frac{\Gamma_R}{\Gamma}
\left( \frac{1}{1 - \frac{\Omega + \mu_R}{\epsilon_F}} -
\frac{1}{1 - \frac{\mu_R}{\epsilon_F}}
\right)
\right]\label{PiRImN}
\end{eqnarray}

Similarly, to ${\cal O}(1/N)$, $\Pi^K$ is given by
\begin{eqnarray}
&&\Pi^K(\Omega) = \frac{-2iN\Gamma \tilde{\Gamma}}{\pi \epsilon_F} \left[\frac{\Gamma_L^2}{\Gamma^2}
sgn(\Omega)\left( \frac{1}{1-\frac{\Omega + \mu_L}{\epsilon_F}}
- \frac{1}{1-\frac{\mu_L}{\epsilon_F}}\right) + \frac{\Gamma_R^2}{\Gamma^2}
sgn(\Omega)\left( \frac{1}{1-\frac{\Omega + \mu_R}{\epsilon_F}}
- \frac{1}{1-\frac{\mu_R}{\epsilon_F}}\right) \right. \nonumber \\
&&\left. + \frac{\Gamma_L \Gamma_R}{\Gamma^2}sgn(\Omega + \mu_R - \mu_L)
\left(\frac{1}{1-\frac{\Omega + \mu_R}{\epsilon_F}}
- \frac{1}{1-\frac{\mu_L}{\epsilon_F}}\right)
+\frac{\Gamma_L \Gamma_R}{\Gamma^2}sgn(\Omega + \mu_L - \mu_R)
\left(\frac{1}{1-\frac{\Omega + \mu_L}{\epsilon_F}}
- \frac{1}{1-\frac{\mu_R}{\epsilon_F}} \right)
\right]
\label{PiKN}
\end{eqnarray}

The bosonic propagators may be evaluated from the Dyson equation
\begin{equation}
D^{-1}_{b,b^*} = D_0^{-1} - \Pi -\delta \Pi^{(1)}
\end{equation}
To ${\cal O}(1/N)$, $\delta \Pi^{(1)} $ does not contribute. So we have
\begin{eqnarray}
D^{R/A,-1}_{b,b^*} = D_0^{-1}-\Pi^{R/A}\\
D^K_{b,b^*} = D^R_{b,b^*} \Pi^K D^A_{b,b^*} \label{DKdef1}
\end{eqnarray}
Evaluating the above, we get
\begin{eqnarray}
Re\left[ D^R_{b,b^*}(\Omega)\right] =
\frac{1}{\Omega - \epsilon_F + E_0 -\frac{N\Gamma}{\pi} \left[ \frac{\Gamma_L}{\Gamma}
\ln\frac{|\Omega + \mu_L - \epsilon_F|}{D}
+ \frac{\Gamma_R}{\Gamma}\ln\frac{|\Omega + \mu_R - \epsilon_F|}{D}   \right] }
\end{eqnarray}
Using the saddle point equation Eq.~\ref{sp22a}, the above becomes,
\begin{eqnarray}
Re\left[ D^R_{b,b^*}(\Omega)\right] = \frac{1}{\Omega -\frac{N\Gamma}{\pi} \left[ \frac{\Gamma_L}{\Gamma}
\ln|\frac{\Omega + \mu_L - \epsilon_F}{\mu_L -\epsilon_F}|
+ \frac{\Gamma_R}{\Gamma}\ln|\frac{\Omega + \mu_R - \epsilon_F}{\mu_R - \epsilon_F}|   \right] }
\label{ReDR}
\end{eqnarray}
Similarly, the imaginary part to {\cal O}(1/N) (where $D^R = Re[D^R] + i Im[D^R]$) is
\begin{eqnarray}
Im\left[ D^R_{b,b^*}(\Omega)\right] = \frac{\frac{-N\Gamma\tilde{\Gamma}}{\pi \epsilon_F}
\left[\frac{\Gamma_L}{\Gamma}\left(\frac{1}{1 - \frac{\Omega + \mu_L}{\epsilon_F}} -
\frac{1}{1 - \frac{\mu_L}{\epsilon_F}}
\right) + \frac{\Gamma_R}{\Gamma}
\left( \frac{1}{1 - \frac{\Omega + \mu_R}{\epsilon_F}} -
\frac{1}{1 - \frac{\mu_R}{\epsilon_F}}
\right)
\right]}{\left(\Omega -\frac{N\Gamma}{\pi} \left[ \frac{\Gamma_L}{\Gamma}
\ln|\frac{\Omega + \mu_L - \epsilon_F}{\mu_L -\epsilon_F}|
+ \frac{\Gamma_R}{\Gamma}\ln|\frac{\Omega + \mu_R - \epsilon_F}{\mu_R - \epsilon_F}|   \right] \right)^2}
\label{ImDR}
\end{eqnarray}

In the evaluation of the spectral-density, we also need the anomalous boson propagators
$D^{R,A,K}_{b,b}, D^{R,A,K}_{b^*,b^*}$. To ${\cal O}\left(\frac{1}{N}\right)$, we find
\begin{eqnarray}
D^R_{b,b}\left(\Omega\right) = D^R_{b^*,b^*}\left(\Omega\right) 
= 2 D^R_{b,b^*}\left(\Omega\right) \delta\Pi^{R(2)}(\Omega)
D^R_{b,b^*}\left(-\Omega\right) \label{DRbb}\\
D^A_{b,b}\left(\Omega\right) = D^A_{b^*,b^*}\left(\Omega\right) 
= 2 D^A_{b,b^*}\left(\Omega\right) \delta\Pi^{A(2)}(\Omega)
D^A_{b,b^*}\left(-\Omega\right)\label{DAbb} \\
D^K_{b,b}\left(\Omega\right) = D^K_{b^*,b^*}\left(\Omega\right) 
= 2 D^R_{b,b^*}\left(\Omega\right) \delta\Pi^{K(2)}(\Omega)
D^A_{b,b^*}\left(-\Omega\right)\label{DKbb}
\end{eqnarray}
with $\delta\Pi^{R,A,K(2)}$ defined in Eqns.~\ref{PiRbb},~\ref{PiAbb} and ~\ref{PiKbb}.

\section{Evaluation of Fermionic self-energy}

\begin{figure}
\includegraphics[totalheight=3cm,width=4cm]{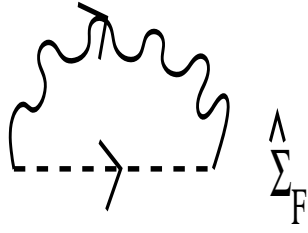}
\caption{Fermionic self-energy}
\label{fig5}
\end{figure}
These are defined as
\begin{eqnarray}
\Sigma^{R,A}_F(1,2) = \frac{i}{2}\left[ D^{R,A}_{b,b^*}(1,2) \Sigma^K_c(1,2) + D^K_{b,b^*}(1,2)
\Sigma^{R,A}_c(1,2)\right]
\label{SigF1}\\
\Sigma^K_F(1,2) = \frac{i}{2}\left[ D^{K}_{b,b^*}(1,2) \Sigma^K_c(1,2) + D^R_{b,b^*}(1,2)\Sigma^{R}_c(1,2) +
D^A_{b,b^*}(1,2)\Sigma^{A}_c(1,2)\right]\label{SigF2}
\end{eqnarray}
and are represented by the diagram in Fig~\ref{fig5}.

\section{Expansion coefficients in the expression for the susceptibility and the conductance} \label{Csi}

The expansion coefficients in Eq.~\ref{neqsusc} are given by
\begin{eqnarray}
C_{S0} &&= m\left(-L_2^{eq} - L_1^{eq} + 2 L_3^{eq} - m J_0^{eq}\right)\\
C_{S1} &&=  -6m L^{eq}_3 +6m L^{eq}_4               
       -m^2 \int_0^{D/T_A} \frac{dx}{\left(x+m \ln\left(1+x\right)\right)^2} \frac{x^2(x+6)}{(1+x)^4}  
       +m^3 \int_0^{D/T_A} \frac{dx}{\left(x+m \ln\left(1+x\right)\right)^3} \frac{2x^3}{(1+x)^4} \\
C_{S2} &&= m L^{eq}_2 -3m L^{eq}_4 -7m L^{eq}_3 -3m L^{eq}_1 +12 m L^{eq}_5              \\  
        &&-m^2 \int_0^{D/T_A} \frac{dx}{\left(x+m \ln\left(1+x\right)\right)^2} 
\frac{x^2(x^2+11x+20)}{2(1+x)^5}        
           +m^3 \int_0^{D/T_A} \frac{dx}{\left(x+m \ln\left(1+x\right)\right)^3} 
\frac{x^3(x+2)}{(1+x)^5}\\
C_{S3} &&= 10 m L_3^{eq} + 3 m L_1^{eq} - 12 m L_4^{eq} - m L_2^{eq} \nonumber \\
&&+m^2 \int_0^{D/T_A} \frac{dx}{\left(x+m \ln\left(1+x\right)\right)^2}\frac{x^2\left(3x-5\right)}
{(1+x)^5} 
+m^3 \int_0^{D/T_A} \frac{dx}{\left(x+m \ln\left(1+x\right)\right)^3}\frac{x^3\left(x+9\right)}
{(1+x)^5}\\
&&-m^4 \int_0^{D/T_A} \frac{dx}{\left(x+m \ln\left(1+x\right)\right)^4}\frac{3x^4}
{(1+x)^5} 
\end{eqnarray}
Whereas the coefficient $C_{G1}$ appearing in the expression for the conductance in Eq.~\ref{GneqNK} is
\begin{eqnarray}
C_{G1} = -2 m L_2^{eq} + 8 m L_3^{eq} - 6 m L_4^{eq} - m^2 \int_0^{D/T_A}
\frac{dx}{\left(x+m \ln\left(1+x\right)\right)^2} 
\frac{2x^3}{(1+x)^4}  
\label{CG1}
\end{eqnarray}
\section{Computation of the spectral-density} \label{speccal}

In this section we give explicit expressions for each of the five diagrams that contribute to
the spectral density,
\begin{eqnarray}
Im{\left[G^R_{f_m,b}\right]} = T_a + T_b + T_c + T_d + T_e   
\end{eqnarray}
where $T_i$ is the $i$-th diagram in Fig~\ref{fig6}.
We find
\begin{eqnarray}
&&T_a + T_b + T_c  = \\
&&-i \frac{\Gamma}{T_A^2\left(1-\frac{\Omega}{T_A}\right)^2}\frac{1}{\left(1 + m_V\right)^2}
\left[1 + \frac{2}{N}\left(\frac{-m^2}{1 + m_V}L_2 + \frac{m^3}{1 + m_V}k + 
\frac{m^3}{(1 + m_V)^2}L_1 \sum_i \frac{\Gamma_i/\Gamma}{\left(1-\mu_i/T_A\right)^2}\right. \right. \nonumber \\
&& \left. \left. - \frac{m^2}{1+m_V}\frac{L_1}{1 - \Omega/T_A} \right)\right] \\
&&+ \frac{i\pi}{N^2 T_A} \frac{m^3}{\left(1 + m_V\right)^2}
\sum_{\alpha \beta}\frac{\Gamma_{\alpha}\Gamma_{\beta}}{\Gamma^2} \int_{-\left(\mu_{\alpha}
-\mu_{\beta}\right)/T_A}^{D/T_A} dx \left(\frac{1}{(1 + x -\frac{\Omega}{T_A})^2} -\frac{1}
{(1-\frac{\Omega}{T_A})^2}\right) 
\frac{\left(\frac{1}{1+x-\frac{\mu_{\beta}}{T_A}} - \frac{1}{1-\frac{\mu_{\alpha}}{T_A}}\right)}
{\left(x + m \sum_i \frac{\Gamma_i}{\Gamma}\ln\left(1 + \frac{x}{1-\frac{\mu_i}{T_A}}\right)\right)^2}
\nonumber \\
&& -\frac{i\pi}{N^2 T_A} \frac{m^3}{\left(1 + m_V\right)^2}\left[ \sum_{\alpha \beta}
\frac{\Gamma_{\alpha}\Gamma_{\beta}}{\Gamma^2}\int_{(\Omega -\mu_{\alpha})/T_A}^{D/T_A}dx 
\frac{1}{\left(1 + x -\frac{\Omega}{T_A}\right)^2}
\frac{\left(\frac{1}{1+x-\frac{\mu_{\beta}}{T_A}} - \frac{1}{1-\frac{\mu_{\beta}}{T_A}}\right)}
{\left(x + m \sum_i \frac{\Gamma_i}{\Gamma}\ln\left(1 + \frac{x}{1-\frac{\mu_i}{T_A}}\right)\right)^2}\right]
\label{div1}\\
&&-\frac{i\pi}{N^2 T_A} \frac{m^3}{\left(1 + m_V\right)^2}\frac{1}{\left(1 -\frac{\Omega}{T_A}\right)^2}
\left[ \sum_{\alpha \beta}
\frac{\Gamma_{\alpha}\Gamma_{\beta}}{\Gamma^2}
\int_{-(\Omega -\mu_{\alpha})/T_A}^{D/T_A}dx 
\frac{\left(\frac{1}{1+x-\frac{\mu_{\beta}}{T_A}} - \frac{1}{1-\frac{\mu_{\beta}}{T_A}}\right)}
{\left(x + m \sum_i \frac{\Gamma_i}{\Gamma}\ln\left(1 + \frac{x}{1-\frac{\mu_i}{T_A}}\right)\right)^2}\right]
\label{div2}\\
&&-\frac{2i\pi}{N^2 T_A} \frac{m^2}{\left(1+ m_V\right)^2}\frac{1}{\left(1 -\frac{\Omega}{T_A}\right)^3}
\sum_{\alpha} \frac{\Gamma_{\alpha}}{\Gamma} \int^{D/T_A}_{-\left(\Omega -\mu_{\alpha}\right)/T_A} dx
\frac{1}{x + m \sum_i \frac{\Gamma_i}{\Gamma}\ln\left(1 + \frac{x}{1-\frac{\mu_i}{T_A}}\right)} \label{div3}
\end{eqnarray}
Note that the expression ~\ref{div1},~\ref{div2} and ~\ref{div3} are logarithmically divergent. These divergences
will be canceled by terms in diagrams (d) and (e), as we show below. 
In particular
\begin{eqnarray}
&&T_d = \frac{2i\pi}{N^2 T_A} \frac{m^2}{\left(1 + m_V\right)^2} \frac{1}{\left(1-\frac{\Omega}{T_A}\right)}
\sum_{\alpha} \frac{\Gamma_{\alpha}}{\Gamma} \int_{\left(\Omega -\mu_{\alpha}\right)/T_A}^{D/T_A} dx
\frac{1}{\left(1 + x -\frac{\Omega}{T_A}\right)^2} \frac{1}{x + m \sum_i \frac{\Gamma_i}{\Gamma} 
\ln \left(1 + \frac{x}{1 - \frac{\mu_i}{T_A}}\right)} \label{div3a}\\
&& + \frac{2i\pi}{N^2 T_A} \frac{m^3}{\left(1 + m_V\right)^2} \frac{1}{\left(1-\frac{\Omega}{T_A}\right)}
\sum_{\alpha \beta} \frac{\Gamma_{\alpha}\Gamma_{\beta}}{\Gamma^2}
\int_{-(\mu_{\beta}-\mu_{\alpha})/T_A}^{D/T_A} dx \left(\frac{1}{1 + x - \frac{\Omega}{T_A}}\right)
\frac{\left(\frac{1}{1+x-\frac{\mu_{\alpha}}{T_A}} - \frac{1}{1-\frac{\mu_{\beta}}{T_A}}\right)}
{\left(x + m \sum_i \frac{\Gamma_i}{\Gamma}\ln\left(1 + \frac{x}{1-\frac{\mu_i}{T_A}}\right)\right)^2} 
\label{div4}
\end{eqnarray}
Note that divergence in Eq.~\ref{div3a} cancels the divergence in Eq.~\ref{div3}. Moreover, we find   
\begin{eqnarray}
&&T_e= \frac{-2i\pi}{N^2 T_A} \frac{m^3}{\left(1 + m_V\right)^2} \frac{1}{\left(1-\frac{\Omega}{T_A}\right)}
\sum_{\alpha \beta} \frac{\Gamma_{\alpha}\Gamma_{\beta}}{\Gamma^2}
\int_{-(\mu_{\alpha}-\mu_{\beta})/T_A}^{D/T_A} dx  \left(\frac{1}{x\left(1-x-\frac{\Omega}{T_A}\right)}\right) 
\nonumber \\
&&\left[
\frac{\ln\left(1 + \frac{x}{1-\frac{\mu_{\beta}}{T_A}}\right) + \ln\left(1 -\frac{x}{1-\frac{\mu_{\alpha}}{T_A}}\right)}
{\left(x + m\sum_{i}\frac{\Gamma_i}{\Gamma}\ln\left(1 + \frac{x}{1-\frac{\mu_i}{T_A}}\right)\right)
\left(x - m\sum_i\frac{\Gamma_i}{\Gamma}\ln\left(1-\frac{x}{1-\frac{\mu_i}{T_A}}\right)\right)}\right]
\label{div1a}\\
&&+ \frac{2i\pi}{N^2 T_A} \frac{m^3}{\left(1 + m_V\right)^2} \frac{1}{\left(1-\frac{\Omega}{T_A}\right)}
\sum_{\alpha \beta} \frac{\Gamma_{\alpha}\Gamma_{\beta}}{\Gamma^2} \int_{-\left(\Omega-\mu_{\alpha}\right)/T_A}^{D/T_A} dx
\left(\frac{1}{x\left(1-x-\frac{\Omega}{T_A}\right)}\right) \nonumber \\
&&\left[
\frac{\ln\left(1 + \frac{x}{1-\frac{\mu_{\beta}}{T_A}}\right) + \ln\left(1 -\frac{x}{1-\frac{\mu_{\beta}}{T_A}}\right)}
{\left(x + m\sum_{i}\frac{\Gamma_i}{\Gamma}\ln\left(1 + \frac{x}{1-\frac{\mu_i}{T_A}}\right)\right)
\left(x - m\sum_i\frac{\Gamma_i}{\Gamma}\ln\left(1-\frac{x}{1-\frac{\mu_i}{T_A}}\right)\right)}\right] \label{div4a}
\end{eqnarray}
Note that Eq.~\ref{div4a} cancels the divergence in Eq.~\ref{div1} and~\ref{div2}, while the divergence in
Eq.~\ref{div4} is canceled by that in Eq.~\ref{div1a}.

\section{Failure of extrapolation to N=2 for systems in equilibrium} \label{extrap}

Many N-fold degenerate magnetic impurity models 
besides being amenable to $1/N$ perturbative approaches, are also 
exactly solvable by Bethe-Ansatz. However,
a comparison between $1/N$ results and exact solutions are not straightforward
as the two approaches use different cut-off schemes (for a discussion on this see~\cite{Bickers87}). 
Therefore the quantities
that may be easily compared are universal quantities that are independent of the cut-off and $T_K$.
One such quantitity is the Wilson ratio  $R= \frac{\pi^2 k_B^2}{J(J+1)g^2 \mu_B^2}\frac{\chi_S}
{\gamma}$, where $\chi_S$ is the impurity susceptibility and 
$\gamma = -\frac{\partial^2 F}{\partial T^2}$ is the specific heat coefficient. 
The exact solution of the Coqblin-Schrieffer
Hamiltonian gives~\cite{Rasul84}
\begin{eqnarray}
R = \frac{N}{N-1}
\end{eqnarray}
Thus for $N=2$, $R=2$. On the other hand a 
perturbative $1/N$ expansion gives~\cite{Read83} $R = 1 + \frac{1}{N}$. Clearly,
setting $N=2$ in this expression gives the rather incorrect result of $R=1.5$, showing that a 
naive extrapolation of the results of
large N to the case of $N=2$ does not work. Another example is the value of the 
zero bias conductance through a
$N$-fold degenerate level in a quantum dot. The exact answer is 
\begin{eqnarray}
G = N\frac{e^2}{h} \frac{4\Gamma_L \Gamma_R}{(\Gamma_L + \Gamma_R )^2}
\sin^{2}\frac{\pi n_F}{N} \label{Geq1}
\end{eqnarray}
$n_F$ being the average charge on the level
which approaches the value $n_F=1$ in the Kondo limit. Therefore when $\Gamma_L = \Gamma_R$
and $N=2$, the conductance in the Kondo limit reaches the maximum possible
value of $2e^2/h$. For $N \gg 1$,  a $1/N$ 
expansion gives $G = \frac{e^2}{h}\frac{\pi^2}{N}$, where again a naive substitution of
$N=2$ gives the incorrect result of $G = \frac{e^2}{h}\frac{\pi^2}{2}$. 

The extrapolation besides giving incorrect numerical values, often does not
capture the qualitative behavior of the temperature dependence of various observables. As an
example let us consider the conductivity for the infinite-U Anderson model 
for a bulk system (rather than a quantum dot). If $\tau^{-1}(\omega,T) = i Im\left[G_{f,b}\right]$
is the scattering rate due to the impurity, then the conductivity in a bulk geometry is
$\sigma^{bulk} \sim \int d\omega \left(-\frac{\partial f}{\partial \omega}\right)\tau(\omega,T)$,
whereas the conductance in a quantum-dot geometry (as has been considered in this
paper) is $G 
\sim \int d\omega \left(-\frac{\partial f}{\partial \omega}\right)\tau^{-1}(\omega,T)$. 
The exact answer for a $N=2$ bulk system is  
\begin{eqnarray}
\frac{\sigma^{bulk}}{\sigma^{bulk}(T=0)} = 1 + c_T \left(\frac{T}{\tilde{T}_K}\right)^2
\label{sigN2}
\end{eqnarray}
where $c_T$ is a positive coefficient.
On the other hand a $1/N$ result for the temperature dependent conductivity 
for the infinite-U Anderson model is~\cite{Read87}
\begin{eqnarray}
\frac{\sigma^{bulk}}{\sigma^{bulk}(T=0)} = 1 + \pi^2 \left(\frac{T}{T_K}\right)^2
\left[1 - \frac{8}{3N}\right]
\end{eqnarray}
Setting $N=2$ in the above equation gives a qualitatively different result from Eq.~\ref{sigN2}
as it predicts that the conductivity will decrease with temperature (rather than increase).

The above discussion shows that large-N results cannot be used to extrapolate to $N=2$. However,
comparison with exact Bethe-Ansatz solutions~\cite{Bickers87} shows that large-N works well for 
$N \geq 4$.

\end{document}